\documentclass{article}
\usepackage[utf8]{inputenc}
\usepackage[english]{babel}
\usepackage[a4paper, margin=3.0cm]{geometry}
\usepackage{amsmath}
\usepackage{amssymb}
\usepackage{caption}
\usepackage{pdfpages}
\usepackage{comment}

\usepackage{titlesec}
\usepackage{multirow}
\usepackage{adjustbox}
\usepackage{subcaption}
\usepackage{graphicx}
\usepackage{array}
\usepackage{booktabs}
\usepackage{float}
\usepackage{adjustbox}
\usepackage{braket}
\usepackage{placeins}
\usepackage{cite}

\usepackage{jheppub}

\usepackage{tikz}
\usetikzlibrary{arrows.meta}
\usetikzlibrary{positioning}
\usetikzlibrary{decorations}
\usetikzlibrary{decorations.markings}

\usepackage{graphicx} 

\title{Paper baryon}
\author{Dario Panfalone, Lorenzo Verzichelli}
\date{\today}
\renewcommand{\figurename}{Fig.}
\DeclareMathOperator{\Tr}{\mathrm{Tr}}
\DeclareMathOperator{\SU}{\mathrm{SU}}
\DeclareMathOperator{\Z}{\mathbb{Z}}
\newcommand{\real}{{\rm Re\,}}

\newcommand{\eq}{\begin{equation}}   
\newcommand{\en}{\end{equation}}   
\newcommand{\eqa}{\begin{eqnarray}}   
\newcommand{\ena}{\end{eqnarray}}   

\newcommand{\bea}{\begin{eqnarray}}   
\newcommand{\eea}{\end{eqnarray}}

\usepackage{orcidlink}
\begin{document}


\title{The Mass of the Baryon Junction: a lattice computation in $2+1$ dimensions}

\author[a]{Michele Caselle,}
\affiliation[a]{Dipartimento di Fisica,  Universit\'a degli Studi di Torino and INFN, Sezione di Torino, Via Pietro Giuria 1, I-10125 Turin, Italy}
\emailAdd{caselle@to.infn.it}

\author[b]{Nicodemo Magnoli,}
\affiliation[b]{Department of Physics, University of Genoa and INFN, Sezione di Genova, Via Dodecaneso 33, I-16146, Genoa, Italy}
\emailAdd{magnoli@ge.infn.it}

\author[a]{Dario Panfalone}
\emailAdd{dario.panfalone@unito.it}

\author[a]{Lorenzo Verzichelli}
\emailAdd{lorenzo.verzichelli@unito.it}




\setcounter{footnote}{0}
\renewcommand\thefootnote{\mbox{\arabic{footnote}}}

\abstract{We present a systematic study of baryonic flux tubes in $\SU(3)$ Yang--Mills theory in $(2\!+\!1)$ dimensions. A recent next-to-leading–order derivation within the Effective String Theory framework has, for the first time, made explicit the corrections proportional to the mass of the baryon junction $M$, up to order $1/R^2$ (where $R$ is the length of the confining strings), opening the possibility of its non-perturbative determination.  One of the main goals of this paper is, through high precision simulations of the three-point Polyakov loop correlator, to measure for the first time the baryon junction mass. By isolating the predicted $1/R^{2}$ term in the open string channel, we obtain the value $M / \sqrt{\sigma} = 0.1355(36)
$, similar to the phenomenological value which is used to describe hadrons, although our computation was done in $(2\!+\!1)$ dimensions. In addition, studying the high temperature behavior of the baryon, we present a new test of the Svetitsky--Yaffe conjecture for the $\SU(3)$ theory in three dimensions.  Focusing on the high temperature regime, just below the deconfinement transition, we compare our lattice results for Polyakov loop correlators with the quantitative predictions obtained by applying conformal perturbation theory to the three-state Potts model in two dimensions and find excellent agreement.}



    \maketitle
\flushbottom

\bibliographystyle{JHEP}
\captionsetup[figure]{labelfont={bf},labelformat={default},labelsep=period,name={Fig.}}
\captionsetup[table]{labelfont={bf},labelformat={default},labelsep=period,name={Tab.}}

\section{Introduction}

A powerful approach to model confinement in Yang-Mills theories is the  "Effective String Theory" (EST) which describes the confining flux tube joining together a quark-antiquark pair as a thin vibrating string~\cite{Nambu:1974zg,Goto:1971ce,Luscher:1980ac,Luscher:1980fr,Polchinski:1991ax}.  

In the last few years significant progress has been made in this context.  In particular it has been realized that the EST enjoys the so called "low energy universality" \cite{Meyer:2006qx,Luscher:2004ib, Aharony:2009gg, Aharony:2011gb, Dubovsky:2012sh,Billo:2012da, Gliozzi:2012cx, Aharony:2013ipa} which states that, due to the peculiar features of the string action and to the symmetry constraints imposed by the Poincar\'e invariance in the target space, the first few terms of its large distance  expansion are fixed and are thus universal. This universality is indeed observed in high precision results from lattice simulations, as reviewed in refs.\cite{Aharony:2013ipa,Brandt:2016xsp,Caselle:2021eir}).  

A particularly challenging issue in this context is represented by baryons in which, in the case of $\SU(N)$ gauge theories, $N$ strings join together at a vertex point which is usually called the "baryon junction". The extension of EST to baryons would be an important non-trivial test of the EST framework and for this reason it has been the subject of a major effort in the past years. More than twenty years ago the dominant term of the EST correction for baryons was evaluated in \cite{Jahn:2003uz,deForcrand:2005vv,Pfeuffer:2008mz} and shown to be different from the one which appears in the EST description of mesons.
On the numerical side lot of efforts were devoted to the evaluation of the ground state of baryons in different dimensions and to the study of the shape of the flux tubes and of the baryon junction  \cite{Alexandrou:2001ip,Alexandrou:2002sn,Takahashi:2000te,Takahashi:2002bw,Bissey:2006bz,Cardoso:2008sb,Bakry:2011vnk,Bakry:2014gea,Borisenko:2018zzd,Koma:2017hcm,Ma:2022vqf,Dmitrasinovic:2009ma,Huebner:2007yzb,Brambilla:2009cd,Sakumichi:2015rfa}.

In this context, a particularly interesting issue is the determination of the mass of the Baryon junction. This value has important consequences both on the phenomenological side (see for instance  \cite{Karliner:2016zzc}) and on the theoretical side since it gives access to the string interaction vertex \cite{Komargodski:2024swh}. The main problem in evaluating this quantity is that it only appears in the next-to-leading term of the EST expansion and was not accessible in old EST calculations \cite{Jahn:2003uz,deForcrand:2005vv,Pfeuffer:2008mz}.
Recently, in ref. \cite{Komargodski:2024swh} Komargodski and Zhong were able to extend the old EST results to the next-to-leading-order, thus making  explicit, both in the open and in the closed string channels, the corrections proportional to the junction mass. This makes the evaluation of the junction mass accessible using Monte Carlo simulations, and this will be one of the two main goals of the present paper.
The second goal will be the study of the high temperature behavior of the baryon, near the deconfinement transition, but still in the confining phase.  In this regime one can resort to an independent effective description based on the mapping of the model to the three-state Potts model in one dimension less.  The interplay between this effective description and the EST results will allow us to better understand a few non-trivial features of the baryonic ground state near the deconfinement point. 

In this paper we  shall concentrate on the (2+1) dimensional $\SU(3)$ gauge theory, which allows high precision simulations with an accessible cost and shares with the more complex (3+1) model most of the features in which we are interested. Moreover, with this choice, we may use existing knowledge on the scale setting and on the values of the string tension and of the critical temperature of the model \cite{Caselle:2024zoh}.

The paper is organized as follows:
In the first two sections, we shall briefly review some known results on the EST description of Baryons (sect.~\ref{EST}) and  on the three-point function of the three-states Potts model in two dimensions (sect.~\ref{SY})  which we shall use as an effective description of the expected behavior of the Baryonic state at high temperature. Sect.~\ref{LAT_SETUP} will be devoted to a description of our lattice setup and of the simulations, while in sect.~\ref{RESULT} we shall discuss our results and compare them with the theoretical predictions. Sect.~\ref{CONCLUSION} will be devoted to some concluding remarks.

\section{EST predictions for the Baryon}
\label{EST}

The usual setting in which EST predictions are tested is the Polyakov loop correlator $\braket{P^+(0) \, P(R)}$ from which the interquark potential is extracted. In the confining regime of Yang-Mills theories, the Polyakov loops are connected by a thin, vibrating flux tube which is well described by the EST~\cite{Nambu:1974zg,Goto:1971ce,Luscher:1980ac,Luscher:1980fr,Polchinski:1991ax}. Following this intuition and thanks to the fact that EST calculations in the case of Polyakov loop correlators can be performed up to very high orders, in the past few years, a remarkable agreement between EST predictions and Monte Carlo simulations of the interquark potential has been observed (see the reviews: \cite{Aharony:2013ipa, Brandt:2016xsp, Caselle:2021eir}). In fact, this agreement is not by chance, but it is a consequence of the so called "low energy universality" \cite{Meyer:2006qx,Luscher:2004ib, Aharony:2009gg, Aharony:2011gb, Dubovsky:2012sh,Billo:2012da, Gliozzi:2012cx, Aharony:2013ipa} which states that, due to the peculiar features of the string action and to the symmetry constraints imposed by the Poincar\'e invariance in the target space, the first few terms of its large distance  expansion are fixed and are thus universal.  This implies that the EST is much more predictive than typical effective models and that corrections with respect to these universal terms appear only at a very high order \cite{Baffigo:2023rin,Caselle:2024zoh, Athenodorou:2011rx,Caristo:2021tbk}. 

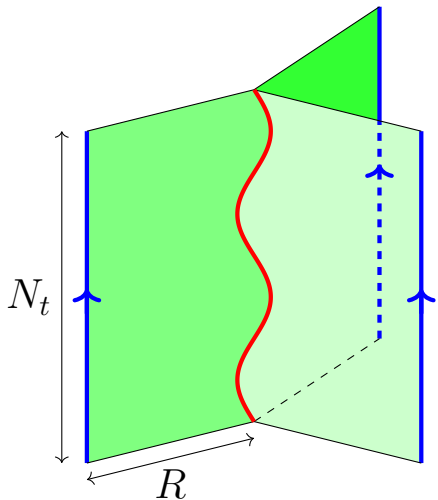
\begin{figure}[!htb]
    \centering

	\tikzset{
		lattice point/.style={
			draw,
			fill,
			circle,
			minimum size=1.0mm,
			inner sep=0,
		},
		intermediate point/.style={
			draw,
			fill,
			circle,
			minimum size=0.0mm,
			white,
			inner sep=0,
		},    
	}
	
	\begin{tikzpicture}[scale=1.1, transform shape]
		
		\pgfmathsetmacro{\xjunction}{3.5}
		\pgfmathsetmacro{\yintercpt}{(0.5 / (5.5-\xjunction) * 0.5 + 3.5)}
		
		
		\fill [green!50] plot [domain=0:6.28, smooth, samples=100, variable=\x]
		({\xjunction - 0.2 * sin(2 * \x r)}, \x / 6.28 * 4)
		-- (1.5, 3.5) -- (1.5, -0.5) -- cycle;
		
		\fill [green!20] plot [domain=0:6.28, smooth, samples=100, variable=\x]
		({\xjunction - 0.2 * sin(2 * \x r)}, \x / 6.28 * 4)
		-- (5.5, 3.5) -- (5.5, -0.5) -- cycle;
		
		\fill [green!80] (5.0, \yintercpt) -- (5.0, 5.0) -- (\xjunction, 4.0);
		
		\draw [blue, ultra thick, ->] (1.5, -0.5) -- (1.5, 1.6);
		\draw [blue, ultra thick] (1.5, 1.3) -- (1.5, 3.5);
		
		\draw [blue, ultra thick, ->] (5.5, -0.5) -- (5.5, 1.6);
		\draw [blue, ultra thick] (5.5, 3.5) -- (5.5, 1.3);
		
		\draw [blue, dashed, ultra thick, ->] (5.0, 1.0) -- (5.0, 3.1);
		\draw [blue, dashed, ultra thick] (5.0, 3.0) -- (5.0, \yintercpt);
		\draw [blue, ultra thick] (5.0, \yintercpt) -- (5.0, 5.0);
		
		\draw (\xjunction, 4.0) -- (5.0, 5.0);
		\draw (\xjunction, 4.0) -- (5.5, 3.5);
		\draw (\xjunction, 4.0) -- (1.5, 3.5);
		
		\draw [thin, dashed] (\xjunction, 0.0) -- (5.0,  1.0);
		\draw (\xjunction, 0.0) -- (5.5, -0.5);
		\draw (\xjunction, 0.0) -- (1.5, -0.5);
		
		%
		
		\draw[domain=0:6.28, smooth, samples=100, variable=\x, ultra thick, red]
		plot ({\xjunction - 0.2 * sin(2 * \x r)}, \x / 6.28 * 4);

    \draw[<->] (1.2, -0.5) -- (1.2, 3.5) node[midway, left] {\Large $N_t$};

\draw[<->] (1.5, -0.7) -- (\xjunction, -0.2) node[pos=0.5, below] {\Large $R$};

	\end{tikzpicture}
    \caption{Graphical representation of the baryon junction. The three blue arrows represent the quarks (or, on the lattice, the Polyakov loops); the worldsheets of the effective string are depicted as surfaces colored in three different shades of green; the wavy red line where the three worlsheets meet illustrates the baryon junction.}
    \label{fig:new1}
\end{figure}

 A non trivial test of the EST framework is its ability to describe also Baryonic configurations in which (in the case of a $SU(N)$ Yang-Mills theory) $N$ confining flux tubes meet at a common point, the \textit{Baryon Junction} (see fig.~\ref{fig:new1}).

This problem was addressed in \cite{Jahn:2003uz,deForcrand:2005vv,Pfeuffer:2008mz,Komargodski:2024swh}
 The main difficulty is that the $N$ strings have  Dirichlet b.c. on the $N$ vertices but non-trivial boundary conditions on the junction. This makes the calculation more complex than in the standard  $\braket{P^+(0)P(R)}$ setting. The first correction beyond the linearly rising term  was obtained more than twenty years ago \cite{Jahn:2003uz,deForcrand:2005vv,Pfeuffer:2008mz} but only recently the next to leading term, encoding the corrections due to the baryon junction, was evaluated in \cite{Komargodski:2024swh} in the particular case of the $\SU(3)$ gauge group and for values of the junction mass $M$ of the same order of $\sqrt{\sigma}$ (which, as we shall see, is correct for the (2+1) dimensional $\SU(3)$ model and is in general expected to be the case for ordinary confining gauge theories) and only for equilateral triangles.  This is exactly the setting in which we shall be interested in the following, thus, we shall briefly review here the results of \cite{Komargodski:2024swh}. Even if our simulations have been performed in the particular case of the (2+1) dimensional $\SU(3)$ model, we shall report here the results for the general $(d+1)$ dimensional case.
 
The result of \cite{Komargodski:2024swh} is\footnote{Notice a slight change of notations with respect to \cite{Komargodski:2024swh}}
\begin{multline}
\braket{P(x_1) P(x_2) P(x_3)} = A(N_t) \frac{e^{-3\sigma R N_t- M N_t}}{\left[\eta(\sqrt{q})\right]^{d}\left[\eta(q)\right]^{d-3}}
\Big(
1+\frac{\pi (d+2)M N_t}{144\sigma R^2} \left[2E_2(q)-E_2(\sqrt{q})\right] \\
+ \mathcal{O}\left( 1 / R^3 \right)
\Big),
\label{eq1}    
\end{multline}
where $x_1,x_2$ and $x_3$ are the three vertices of an equilateral triangle, $R$ denotes the distance of these vertices with respect to the Fermat point (which in the equilateral case coincides with the circumcenter of the triangle), $N_t$ is the size of the lattice in the compactified "time" direction and $\sigma$ is the zero temperature string tension.  $\eta({q})$ is the Dedekind function and  $E_2(q)$ is the Eisenstein series of order 2 with $q=e^{-\frac{\pi N_t}{R}}$  (see Appendix \ref{App:C} for definitions and properties). The overall function $A(N_t)$ is an undetermined term which does not depend on $R$ but contains a non-universal dependence on $N_t$ due to the Dirichlet boundary conditions at the vertices of the triangle.  This is nothing else than the ordinary "perimeter" term that we have in the standard $\braket{P^+(0)P(R)}$ correlator, but in this setting it has a non trivial effect since it interferes with the term proportional to $e^{-M N_t}$ in front of the equation and prevents estimating the junction mass from this term. Thus, the only way to estimate the junction mass is by resorting to the next-to-leading term,  which is proportional to $N_t/R^2$ and is not affected by any contamination from non-universal corrections. 
 
The above expression can be expanded in the two limits of the "open string" channel (i.e. $N_t \gg R$ limit of eq.~\eqref{eq1}) and the "closed string" channel ($R \gg N_t$ limit of eq.~\eqref{eq1}, with $N_t\sqrt{\sigma} \gg 1$).

\subsection{Open string channel}
In this limit eq.~\eqref{eq1}  becomes:

\begin{equation}
\braket{P(x_1) \, P(x_2) \, P(x_3)}    = A_{open} (N_t) \, e^{-N_t E_0(R)},
\label{eqlowT}
\end{equation}
where the prefactor $A_{open} (N_t) =  A (N_t)  e^{-N_t M} $
does not depend on $R$ and all the $R$ dependence is concentrated in the ground 
state energy $E_0(R)$ of the Baryon (the "baryonic potential" in LGT jargon) defined as

 \begin{equation}
        E_{0}(R) =  3R\sigma - \frac{(d-2) \pi}{16 R} - \frac{(d+2)M \pi}{144 \sigma R^2} + \mathcal{O}(1/R^3).
        \label{eq2}
\end{equation}

Notice that, as it happens for the interquark potential, in this limit, there is no term proportional to $\log R$ in $E_0(R)$

The peculiar feature of this expression is that  (as already observed in \cite{Jahn:2003uz,deForcrand:2005vv,Pfeuffer:2008mz})  in $d=2$ the $1/R$ term (the analogous of the "L\"uscher term") vanishes and the first correction is the $1/R^2$ term due to the baryon junction. This makes the $d=2$ case particularly effective to measure the junction mass $M$.

\subsection{Closed string channel}

This limit can be reached by performing a modular transformation of eq.~\eqref{eq1} and then expanding. One finds:
 
 \begin{equation}
 \braket{P(x_1) P(x_2) P(x_3)}    = A_{closed}(N_t)\left(\frac{N_t}{R}\right)^\frac{2d-3}{2} e^{-3 R E_0(N_t)},
\label{eqhighT}
\end{equation}
with
 \begin{equation}
        E_{0}(N_t) =  \sigma N_t-\frac{\pi (2d-3)}{6N_t} + \mathcal{O}\left( 1 / {N_t}^3 \right)
        \label{eq5},
\end{equation}
and
\eq
A_{closed} (N_t) =  A (N_t) \, e^{-MN_t \left[1-\frac{\pi (d-2)}{18\sigma N_t^2} + \mathcal{O}\left( 1 / {N_t}^4 \right)\right]}.
\en
Let us stress a few important features of this result.

First, differently from the open string channel case, in this limit the correlation function  contains an additional $R$ dependent prefactor with an exponent which is fixed by the EST analysis to be $-\frac{2d-3}{2}$. This term can be easily detected in the simulations and will play an important role in the following.

Another important difference is that in the closed string channel expression for $E_0$ of eq.~\eqref{eq5} a "L\"uscher type" term proportional to $1/N_t$ is present\footnote{This is due to the different argument of the two Dedekind functions in eq.~\eqref{eq1}, the exact cancellation of the open string channel is modified by the modular transformation in the $\frac{\pi (d-1)}{2N_t}$ term of eq.~\eqref{eq5}.}.

Moreover, if we fix the value $d=2$ in which we are interested the $R$ dependence of eq.~\eqref{eqhighT} becomes
\eq
\braket{P(x_1) \, P(x_2) \, P(x_3)}  \sim \frac{e^{-3 R E_0(N_t)}}{\sqrt{R}},
\label{eq7}
\en
with
\eq
E_0(N_t)=\sigma N_t\left(1-\frac{\pi}{6\sigma N_t^2} + \mathcal{O}\left( 1 / {N_t}^4 \right) \right).
\label{eq8}
\en

These expressions are remarkably similar to the analogous ones, for the $\langle P^{\scriptscriptstyle+}(0)P(R) \rangle$ correlator in the closed string,
\eq
\braket{P^+(0)P(R)} \sim \frac{e^{-R E_{P^{\scriptscriptstyle+}P}(N_t)}}{\sqrt{R}},
\en
with 
\eq
E_{P^+P}(N_t)=\sigma N_t\sqrt{1-\frac{\pi}{3\sigma N_t^2}}, \footnote{The leading order correction here is due to the deviation from the Nambu--Got\=o action studied in \cite{Caristo:2021tbk, Caselle:2024zoh} which is $\mathcal{O}\left( 1 / {N_t}^7 \right)$.}
\label{eq9}
\en
whose expansion at the first order (the same that we are addressing in the Baryonic case), is exactly
\eq
E_{P^+P}(N_t)\sim \sigma N_t\left(1-\frac{\pi}{6\sigma N_t^2}+ \cdots\right),
\label{eq10}
\en
thus we see that in this limit the two ground state energies coincide.

As usual, these EST results are large distance expansions. In particular, in this setting, they are expansions in powers of 
$\frac{1}{N_t\sqrt{\sigma}}$. The combined constraints $R \gg N_t$ and $N_t\sqrt{\sigma} \gg 1$ make these predictions particularly difficult  to test in the lattice simulations. In the closed string channel one is typically interested in studying the high temperature regime of the theory, near the deconfinement transition but still in the confined phase, where $N_t$ is small and the $R \gg N_t$ condition can be realized with an acceptable numerical effort. However in this limit $N_t\sqrt{\sigma}\sim 1$ and higher order terms of the Nambu--Got\=o action become important. In the present case we have no EST result for these higher order terms, but we can resort to an alternative approximation which we shall discuss in detail in sect.~\ref{SY} below i.e. the possibility to construct an alternative description for the $\braket{P(x_1) P(x_2) P(x_3)}$ based on the mapping, in the vicinity of the deconfinemnet transition, of the $\SU(3)$ gauge model in (2+1) dimensions to the three-states Potts model in two dimensions. We shall see below that the comparison of these two approximate descriptions may allow to guess, at least to some extent, the behavior of the higher order Nambu--Got\=o corrections in the baryonic potential.

\subsection{On the relevance of the Baryon junction mass}

The value of the baryon junction mass $M$ plays an important role in our understanding of EST and more generally of colour confinement \cite{Komargodski:2024swh}.  

In particular, its sign and magnitude plays a crucial role. When $M>0$ the theory describes a regime where long confining strings are weakly coupled, and the configuration is both perturbatively stable and unitary \cite{Komargodski:2024swh}.
In contrast, when $M<0$, the interaction becomes strong. A negative value of $M$ larger in magnitude than $\sqrt{\sigma}$ leads to a perturbative instability and is therefore ruled out. Moreover, if $M$ has a moderate negative value, $-\sqrt{\sigma}\lesssim M< 0$, the strong coupling suggests possible violations of unitarity and thus signals the necessity of further investigation of the theory \cite{Komargodski:2024swh}.

Moreover, in the closed channel, the junction may be interpreted as a localized interaction vertex where three closed strings merge. The mass of the junction provides a quantitative prediction of the strength of this interaction \cite{Komargodski:2024swh}.

From a phenomenological point of view, the mass of baryon junction plays a role in a wide class of models of exotic hadrons, therefore, a lattice determination provides a stringent benchmark for these constructions.

Finally, several holographic models of confining gauge theories also have predictions for the junction mass. Our results could be used to discriminate among these models.

\section{An alternative description of the baryon in the vicinity of the deconfinement transition}
\label{SY}

\subsection{The Svetitsky--Yaffe mapping}
A powerful tool to address the physical behavior of Yang-Mills theories in the vicinity of the deconfinement transition is through the construction
of an effective action for the Polyakov loops only, which can be obtained by integrating out the space-like links of the model.
 Such a construction corresponds in all respects to a "dimensional reduction":
starting from a ($d+1$) dimensional LGT we end up with an effective action for the Polyakov loops which will be a
$ d$-dimensional spin model with global symmetry the center of the original gauge group.

While the explicit construction of such an effective action may be cumbersome and, in general, can be performed only as a strong coupling expansion,
some general insight on the behavior of the model can be deduced by simple renormalization group arguments~\cite{Svetitsky:1982gs} .

Indeed, even if as a result of the integration over the original gauge degrees of freedom we may expect long-range interactions between the Polyakov loops, 
it can be shown~\cite{Svetitsky:1982gs} that these interactions decrease exponentially with the distance. 
Thus, if the phase transition is continuous, in the vicinity of the critical point, the fine details of the interactions can be neglected, and the model will belong to the same universality class of the simplest spin model, with only nearest neighbors interactions, sharing the same symmetry-breaking pattern.  
This is exactly our case, since both the deconfinement transition of the $\SU(3)$ LGT in (2+1)-dimensions and the magnetization transition of the three-states Potts model in two dimensions are continuous. They thus  belong to the same universality class.

This mapping has several important consequences:
\begin{itemize}
\item[{\em a)}]
The ordered (low temperature) phase of the spin model corresponds to the 
deconfined (high temperature) phase of the original gauge theory. This is the phase in which both the Polyakov loop, in the original LGT, and the spin, in the effective spin model, acquire a non-zero expectation value.  
\item[{\em b)}]
The Polyakov loop is mapped into the spin operator while the plaquette is mapped into the energy operator of the effective spin model. Accordingly, the Polyakov loop correlators in the confining phase in which we are interested are mapped into the spin-spin correlator of the disordered, high temperature phase of the spin model.

\item[{\em c)}]
Thermal perturbations from the critical point in the original gauge theory, which are driven by the plaquette operator, are mapped into thermal perturbation of the effective spin model which are driven by the energy operator. Notice however the change in sign: an {\sl increase} in temperature of the original gauge theory corresponds to a {\sl decrease} of the temperature of the effective spin model. 

\end{itemize}
We cannot exclude the presence in the Yang--Mills theory of other operators (beyond the plaquette) that deform the theory. This would possibly produce non universal contributions to the correlators different from those in the Potts model. We will stress in sect.~\ref{results_SY}, where we present the fit model, which terms are universal (and can be deduced from the Potts model) and which are not (and must be fitted). We will also show how the deviation from the Potts model are compatible with zero, with our data.

A major consequence of this correspondence is that, in the vicinity of the deconfinement point, the behavior of the interquark potential is strongly constrained and thus it represents, as we shall see, a powerful tool to test the predictions of the effective string model. In particular for the three-states Potts model, which is the case of interest for us, thanks to the exact integrability of the model, we have a very good analytical control of the large distance behavior of both  the spin-spin and the three spins correlators \cite{Caselle:2005sf} and thanks to conformal perturbation we have a complementary short distance expansion for both correlators.
We report here the main results. Details on the derivation can be found in the Appendix A and in the original papers \cite{Caselle:2005sf,Guida:1995kc}.

\subsection{Two-point correlator in the spin model}
Representing each spin as phase $s = e^{2 i \pi \, n / 3}$, where $n$ is an integer modulo three, the leading contribution in the large distance expansion of the two-point function (\textit{i.e.} the spin-spin correlator) reads:
\eq
\braket{s^+(0)s(R)} \ \sim \  K_0(mR)
\label{eq11}.
\en
Where $K_0(x)$ denotes the modified Bessel function of order zero and $m$ is the mass scale of the model (i.e., the inverse of the correlation length).

The same two-point function, also admits a short-distance expansion of the following form:
\eq
\braket{s^+(0)s(R)} \ \sim \ \frac{1}{r^{4/15}} \left(1+g_1 r^{4/5}+g_2 r^{6/5}+g_3 r^{2}+g_4 r^{12/5}+ \dots \right),
\en
where $r=mR$ and the functional form of the expansion is fixed by the scaling dimensions of the operators of the three-state Potts model. The first two coefficients $g_1$ and $g_2$  can be evaluated exactly using conformal perturbation and are given by: 
\begin{equation}
    g_1=-1.59936 \dots\;, \ \ \ g_2=0.805622 \dots\;,
\end{equation}
while the next two coefficients were estimated numerically in \cite{Caselle:2005sf} and are:
\begin{equation}
    g_3 = 0.167(18) \ , \ \  g_4 = -0.130(23).
\end{equation}
These coefficients are almost universal, the only non-universal part is represented by the Vacuum Expectation Values (VEV) of the various operators in the perturbed theory (see \cite{Caselle:2005sf}) and Appendix \ref{App:SY} for further details).


\subsection{Three-point correlator in the spin model}
Likewise the two-point function, also the three-point function (\textit{i.e.} the correlator of three spins) admits a large distance expansion. In case of triangles that are close to being equilateral, the leading contribution is the following:
\eq
\braket{s(x_1)s(x_2)s(x_3)} \ \sim \  K_0(m R_Y),
\label{eq12}
\en
where it is assumed that all the angles of the triangle $(x_1,x_2,x_3)$ are less than $2 \pi /3$ and   
$R_Y$ denotes the minimal total length of lines connecting the 3 spins to a single point (the Fermat Point) and is given by   
\eq
R_Y \ = \ R_1 + R_2 + R_3 ,
\en   
where $R_i$ is the length of the segment joining the vertex $i$ with the Fermat point.
For equilateral triangles, all the $R_i$ are equal and $R_Y=3R$.  

Finally, we report the short-distance expansion of the three-point function. In this case, we limit ourselves only to the expression for the equilateral triangle:
\eq
\braket{s(x_1)s(x_2)s(x_3)} \ \sim \  \frac{1}{r^{2/5}} \left( 
c_1 + c_2 \, r^{4/5} + c_3 \, r^{6/5}  +c_4 \, r^2 + c_5 \, r^{12/5}  + \dots   
\right)  .
\label{eqsd}
\en

where, now $r$ is the side of the triangle in units of the mass scale of the model: $r = \sqrt{3} R \, m$ and the exponents are fixed by the scaling dimensions of the various operators.  The first two coefficients can be estimated analytically and are:
\begin{equation}
    c_1 = 1.09236\dots \;,\ \ \ \  c_2 = -2.29795\dots,
\end{equation}
while the last three were estimated numerically in \cite{Caselle:2005sf} to be:
\begin{equation}
    c_3 = 1.24(10) \ , \ \   c_4 = 0.44(45) \ , \ \    c_5 = -0.33(37).
\end{equation}
\subsection{Comparison with EST results}
\label{comparison}
As mentioned above
these expressions must agree with those obtained with the EST approach in the closed string channel limit. This comparison has already been performed in \cite{Caselle:2024zoh} for the two-point function.  We are now in the positon to extend it also to the much more complex case of the three-point function. Indeed, if we recall the large distance expansion of the $K_0$ Bessel function:
\begin{equation}
    K_0(x)\sim \sqrt{\frac{\pi}{2x}} \, e^{-x},
\end{equation}
we immediately see that eq.~\eqref{eq12} agrees with the EST expression of eq.~\eqref{eq7} if we identify the mass $m$ of the Potts model with the ground state energy $E_0(N_t)$ of eq.~\eqref{eq8}. A highly non-trivial test is then represented by the argument of these asymptotic expansions: looking at eq.s \eqref{eq11} and \eqref{eq12} we see that the two and three-point functions must share the same mass scale: $m$. Looking at eq.s \eqref{eq8} and \eqref{eq10}, we see that exactly the same pattern is reproduced in the EST calculation if we truncate the EST prediction for the two-point function to the same order at which the Baryonic calculation is performed. It is rather impressive to notice that this agreement is the result of a set of nontrivial cancellations among different Dedekind functions. 
 
From the comparison between the two effective descriptions, we may extract a few important consequences for the EST model.

First, the fact the mass scale governing the two-point and the three-point functions must coincide suggests that, at least near the deconfinement transition, the higher order corrections to the three quarks potential can be resummed exactly as one does (thanks to the exact solution of the Nambu--Got\=o EST for the $\braket{P^+P}$ correlator \cite{Luscher:2004ib, Billo:2005iv}) in the case of the interquark potential. This suggests to consider the result of eq.~\eqref{eq8} only as the first order term of a large distance expansion, and resum it as follows:
\eq
E_0(N_t)=\sigma N_t\left(1-\frac{\pi}{6\sigma N_t^2}\right) \hskip 0.3cm \longrightarrow \hskip 0.3cm
E_0(N_t)=\sigma N_t\sqrt{1-\frac{\pi}{3\sigma N_t^2}}.
\label{eq8bis}
\en
This resummation makes essentially no difference for intermediate and low values of the temperature, but near the deconfinement transition, the effect becomes important. Let us stress, however, that this resummation is only inferred from the Svetitsky--Yaffe mapping and the knowledge of the two point function. Thus we expect it to receive corrections close to the deconfinement temperature, similarly to what   happens for the two point function ground state (see ref.~\cite{Caselle:2024zoh}).

Second, as we know very well by now, the Nambu--Got\=o theory is only the first-order term of the large-distance expansion of the EST. As $R$ decreases, higher order corrections become important. The short distance expansion of eq.~\eqref{eqsd} allows to have an independent look at these higher order terms. A non-trivial test of the fact that we are looking to the same EST expression from different points of view is that the ground state energy extracted in the two limites (i.e. using eq.~\eqref{eq12} and eq.~\eqref{eqsd} must coincide.

In what follows, we shall use our Montecarlo results to test both these observations.

\section{Lattice Setup}
\label{LAT_SETUP}

We used the same numerical setup of ref.~\cite{Caselle:2024zoh}, which we briefly review in this section.

We regularize the $\SU(3)$ Yang--Mills theory on a cubic lattice with lattice spacing $a$, with periodic boundary conditions in the three directions and volume $V=(N_s^2 \times N_t)$. Note that we intend $N_t$ and $N_s$ to be dimensionful, so the actual number of lattice points in the time direction is $N_t / a$ and $N_s / a$ likewise in the space direction. We discretize the purely gluonic action using the standard Wilson plaquette action:
\begin{equation}
  S_W [U] = \beta \sum_{x, \mu < \nu} \left( 1 - \frac{1}{3} \real \Tr \Pi_{\mu \nu} (x) \right) ,
\end{equation}
where $\beta$ is the Wilson parameter defined in terms of the lattice spacing $a$ and the bare gauge coupling $g$ as $\beta=6/(ag^2)$ and $\Pi_{\mu\nu}(x) = U_\mu(x) U_\nu(x+\hat{\mu}) U_\mu^\dagger(x+\hat{\nu}) U_\nu^\dagger(x)$  is the plaquette operator on the site $x$ along the $(\mu, \nu)$ plane.
In this formulation, at fixed $\beta$, the temperature is equivalent to the inverse of the extent of the system in the Euclidean time direction $T = 1/N_t$.

As it is well known, this theory exhibits a finite-temperature phase transition associated with the spontaneous breaking of the $\Z_3$ center symmetry. The order parameter is the expectation value of the Polyakov loop, which transforms non-trivially under the symmetry group, defined as:
\begin{equation}
    P(\Vec{x}) = \frac{1}{3} \Tr \left[ \prod_{t = 0}^{N_t} U_0\left(\Vec{x}, t \right) \right].
\end{equation}
When $T < T_c$ the  expectation value of the Polyakov loop remains zero, reflecting the unbroken $\Z_3$ center symmetry. Conversely, when $T>T_c$ the theory is in the deconfined phase, where the spontaneous breaking of this symmetry gives rise to a non-zero expectation value for the Polyakov loop.

In this work, we study the system in the confining phase, both in the vicinity of the deconfinement transition ($0.75<T/T_c<0.9$) and in the low temperature regime ($T/T_c\simeq0.18$).

We investigate the three-point correlation function of Polyakov loops as a function of their relative distance
\begin{equation}
    G^{(3)}(x_1,x_2,x_3) = \frac{a^2}{N_s^2} \left\langle \sum_{\Vec{x_1}\Vec{x_2}\Vec{x_3}} P(\Vec{x_1}) P(\Vec{x_2} )P(\Vec{x_3} ) \right\rangle.
\end{equation}
Ideally, our goal is to measure the correlation function at equidistant points corresponding to the vertices of an ideal equilateral triangle. However, due to the inherent constraints imposed by the cubic structure of the lattice, it is not feasible to perfectly realize such an arrangement. Therefore, we compute the observable using isosceles triangles that approximate the equilateral configuration as closely as possible. In practice, we choose the base of the triangle $b$ to be an even integer in units of the lattice spacing $a$. Then the height of the isosceles triangle (again in units of $a$) will be the closest integer to $\sqrt{3} \, b / 2$. For $b \ge 6 a$, this prescription always allows to construct triangles whose equal sides differ from the base by less than $3 \%$.

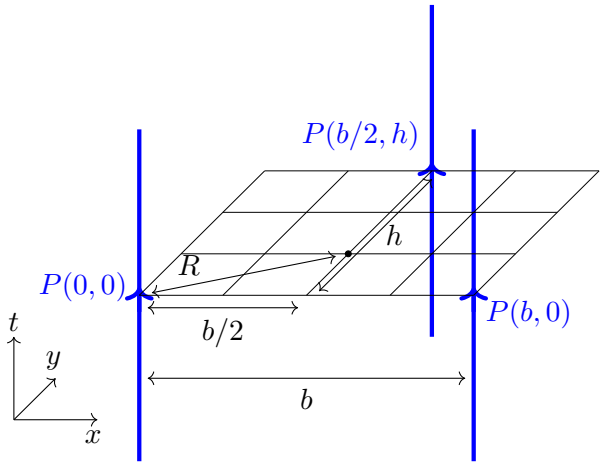
\begin{figure}[!htb]
    \centering

	\tikzset{
		lattice point/.style={
			draw,
			fill,
			circle,
			minimum size=1.0mm,
			inner sep=0,
		},
		intermediate point/.style={
			draw,
			fill,
			circle,
			minimum size=0.0mm,
			white,
			inner sep=0,
		},    
	}
	
	\begin{tikzpicture}[scale=1.1, transform shape]
		
		\draw [ -> ] (0, 0) -- (0, 1) node [pos=0.95, anchor=south] {$t$};
		\draw [ -> ] (0, 0) -- (1, 0) node [pos=0.95, anchor=north] {$x$};
		\draw [ -> ] (0, 0) -- (0.5, 0.5) node [pos=0.95, anchor=south] {$y$};
		
		\draw [blue, ultra thick, ->] (1.5, -0.5) -- (1.5, 1.6) node [anchor=east] {$P(0, 0)$};
		\draw [blue, ultra thick] (1.5, 1.3) -- (1.5, 3.5);
		
		\draw [blue, ultra thick, ->] (5.5, -0.5) -- (5.5, 1.6) node [anchor=north west] {$P(b, 0)$};
		\draw [blue, ultra thick] (5.5, 3.5) -- (5.5, 1.3);
		
		\draw [blue, ultra thick, ->] (5.0, 1.0) -- (5.0, 3.1)
		node [anchor=south east] {$P(b/2, h)$};
		\draw [blue, ultra thick] (5.0, 3.0) -- (5.0, 5.0);
		
		\draw [ultra thin] (1.5, 1.5) -- (5.5, 1.5);
		\draw [ultra thin] (2.0, 2.0) -- (6.0, 2.0);
		\draw [ultra thin] (2.5, 2.5) -- (6.5, 2.5);
		\draw [ultra thin] (3.0, 3.0) -- (7.0, 3.0);
		
		\draw [ultra thin] (1.5, 1.5) -- (3.0, 3.0);
		\draw [ultra thin] (2.5, 1.5) -- (4.0, 3.0);
		\draw [ultra thin] (3.5, 1.5) -- (5.0, 3.0);
		\draw [ultra thin] (4.5, 1.5) -- (6.0, 3.0);
		\draw [ultra thin] (5.5, 1.5) -- (7.0, 3.0);
		
		\filldraw (4.0, 2.0) circle (1 pt);
		
		
		\draw [ <-> ] (1.6, 0.5) -- (5.4, 0.5) node [pos=0.5, anchor=north] {$b$};
		\draw [ <-> ] (1.6, 1.35) -- (3.4, 1.35) node [pos=0.5, anchor=north] {$b / 2$};
		\draw [ <-> ] (3.65, 1.55) -- (5.0, 2.9) node [pos=0.5, anchor=west] {$h$};
		\draw [ <-> ] (1.65, 1.53) -- (3.85, 1.97) node [pos=0.2, anchor=south] {$R$};
		
	\end{tikzpicture}
    \caption{Geometry of the three-point function on the lattice.}

    \label{fig:new2}
\end{figure}

Following this construction, we define in the following the distance $R$ of each Polyakov loop from the baryon junction as
\begin{equation}
    R = \frac{R_Y}{3} = \frac{b}{2 \, \sqrt{3}} + \frac{h}{3},
    \label{eqRY}
\end{equation}
where $h$ is the height of the isosceles triangle which best approximates the equilateral geometry (see fig.~\ref{fig:new2})
and $R_Y$ is the total length of the three flux tubes.

We refer the interested reader to the Appendix \ref{App:B} for a detailed discussion of this choice and its comparison with other possible choices.

We carefully analyzed the deviations from the ideal equilateral triangle and assessed their impact on the extracted physical observables (see Appendix \ref{App:B}). We verified that the systematic uncertainties introduced by the lattice geometry are minimized and properly accounted for in our results.

A major advantage of studying the $\SU(3)$ model in $(2+1)$ dimensions is that we may leverage previous studies to fix the parameters of the model.  In particular, a precise determination of $\beta_c$ at different values of $N_t$ based on results from ref.~\cite{Liddle:2008kk} is given by the following formula~\cite{Caselle:2011fy} valid for generic values of $N$:
\begin{align}
\label{eq:scale_setting}
\frac{T}{T_c} = \frac{a}{N_t} \frac{\beta - 0.22N^2 + 0.5}{0.375N^2+0.13-0.211/N^2}.
\end{align}

We performed our simulations at four different values of $\beta$. Using eq.~\eqref{eq:scale_setting}, we tuned the values of $\beta$ to fix the lattice spacing to four different values, namely $a=1/(10.5 T_c)$, $a=1/(11.5 T_c)$, $a=1/(12.5 T_c)$ and $a=1/(13.5 T_c)$. The first of these values, which corresponds to $\beta=36.33$ is in common with those studied in \cite{Caselle:2024zoh} and allowed us to benchmark our results, the other three correspond to finer lattice spacings, a choice which turned out to be mandatory to study with enough precision the three-point function. 
We chose the temporal extent $N_t$ so as to keep the temperature fixed at $T/T_c=0.18-0.19$, or in the range ($0.75<T/T_c<0.9$). 
\begin{table}[!htb]
\centering
\begin{tabular}{|l|l|l|l|l|l|}
\hline
$\beta$ & $N_t / a$ & $N_s / a$ & $T/T_c$ & $n_{conf}$ & $n_{update}$ \\ \hline
36.33   &  56     &  72     &   0.19      &  $1.9\times10^3 $      &  500                       \\ \hline
39.65   &  63     &  72     &   0.18      &  $8\times10^2 $        &  1000                       \\ \hline
42.97   &  70     &  72     &   0.18     &   $5\times10^2 $        &  1000                      \\ \hline
46,29   &  77     &  80     &   0.18      & $ 5\times10^2 $        &  1000                    \\ \hline
\end{tabular}
\caption{Parameters of our simulations of the $\SU(3)$ Yang--Mills theory, for different values of the lattice spacing $a$. The gauge configurations were updated using a multilevel Lüscherr-Weisz algorithm \cite{Luscher:2001up}, implemented with one inner level. $n_{conf}$ refers to the number of configurations used in the simulations, while $n_{update}$ indicates the number of updates performed at the inner level.} 
\label{tab:lowTsimdetails}
\end{table}

\begin{table}[!htb]
\centering
\begin{tabular}{|l|l|l|l|l|}
\hline
$\beta$                & $N_t / a$ & $N_s / a$ & $T/T_c$ & $n_{conf}$        \\ \hline
36.33                  & 12    & 120   & 0.87    & $2.9 \times 10^5$ \\ \hline
\multirow{2}{*}{39.65} & 13    & 160   & 0.88    & $1.7 \times 10^5$ \\ \cline{2-5} 
                       & 14    & 160   & 0.82    & $1.9 \times 10^5$ \\ \hline
\multirow{3}{*}{42.97} & 14    & 160   & 0.89    & $1.9 \times 10^5$ \\ \cline{2-5} 
                       & 15    & 160   & 0.83    & $1.9 \times 10^5$ \\ \cline{2-5} 
                       & 16    & 160   & 0.78    & $1.8 \times 10^5$ \\ \hline
\multirow{4}{*}{46.29} & 15    & 200   & 0.90    & $1.9 \times 10^5$ \\ \cline{2-5} 
                       & 16    & 200   & 0.84    & $1.6\times 10^5$  \\ \cline{2-5} 
                       & 17    & 200   & 0.79    & $1.4 \times 10^5$ \\ \cline{2-5} 
                       & 18    & 200   & 0.75    & $1.5 \times 10^5$ \\ \hline
\end{tabular}
\caption{Information on the simulations at temperature $T$ close to the deconfinement critical temperature $T_c$, for different value of the lattice spacing $a$. }
\label{tab:simdetSY}
\end{table}

At each value of $\beta$ we computed the three-point Polyakov loop correlator $G^{(3)}(R)$ for various values of the distance  among quarks, we measured correlators at different distances on independent configurations, ensuring that each measurement is uncorrelated with others.
Further information on the simulations is reported in tab.~\ref{tab:lowTsimdetails} for the low $T$ simulations and in tab.~\ref{tab:simdetSY} for those at high $T$.

Note that, at low temperature, the signal of the three-point function significantly drops even at short space separation, due to the large temporal extent. However, the implementation of L\"uscher-Weisz "multilevel" algorithm \cite{Luscher:2001up} with one internal level allows us to effectively fight the exponential decay of the signal-to-noise ratio.
In particular, for $T = 0.18-0.19 T_c$, we divided the lattice into seven slabs and updated each one $\mathcal{O}(10^3)$ times independently. The exact number of updates at each value of $\beta$ is reported in the last column of tab.~\ref{tab:lowTsimdetails}.

\section{Numerical results}
\label{RESULT}

\subsection{Analysis of the low $T$ data}
\subsubsection{Baryon Junction Mass}

Following the discussion of sect.~\ref{EST}
the Baryon Junction mass $M$ can be estimated looking at the $1/R^2$ correction of the ground state energy $E_{0}(R)$ in the open string channel. 
For each value of $\beta$ and $N_t$ we fitted our data for  $G^{(3)}(R)$ with the prediction of eq.~\eqref{eqlowT} which we report here for completeness:
\begin{equation}\label{eq:threeptfitfunc}
    G^{(3)}(R) = A_3 \, e^{-N_t \Big(3 R \sigma - \frac{M \pi}{36 \sigma R^2} \Big)},
\end{equation}
where $A_3$, $\sigma$ and $M$ are the fit parameters.

\begin{figure}[!htb]
    \centering
    \includegraphics[width=\linewidth]{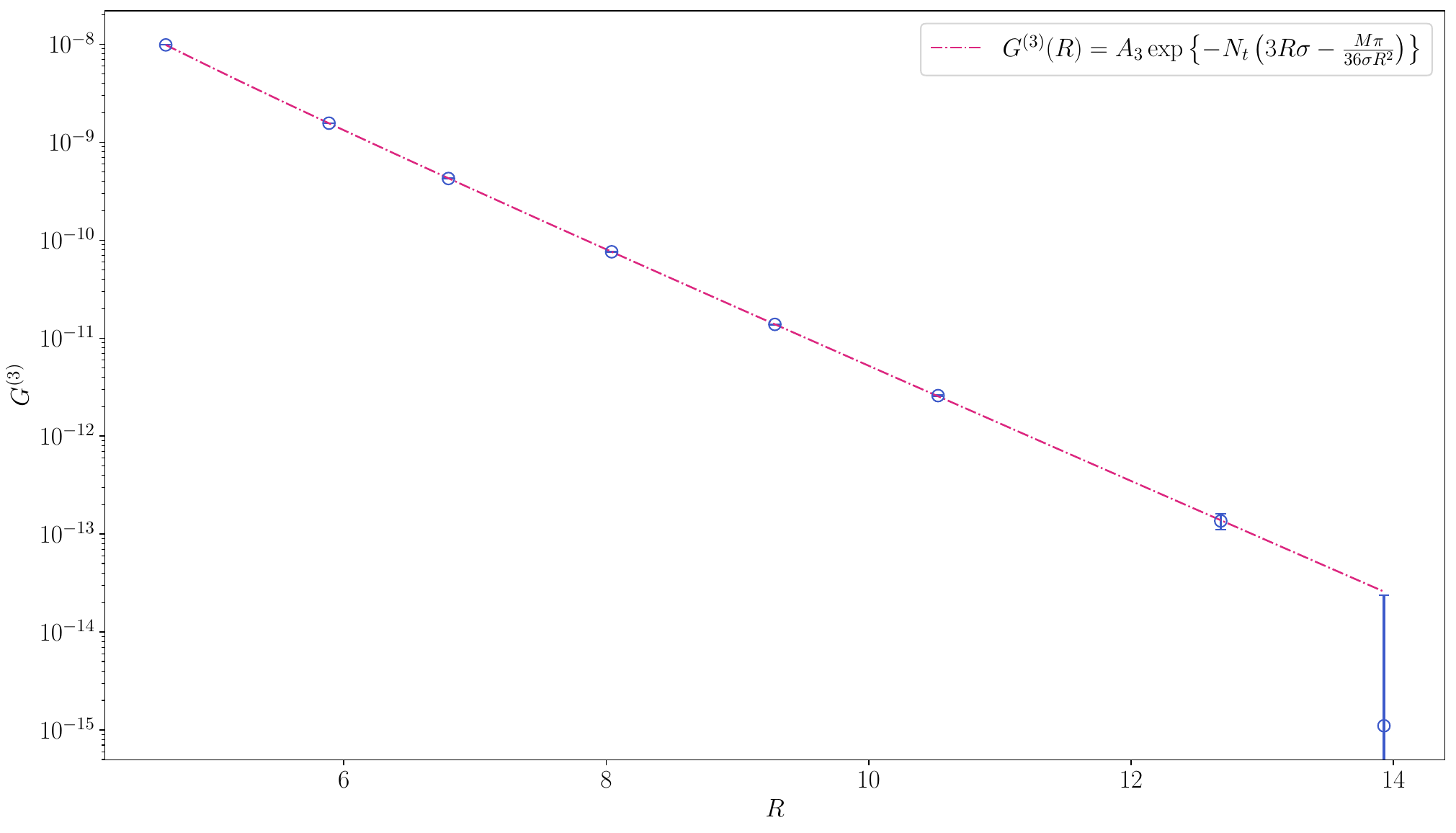}
    \caption{Best fit of the data at $\beta=42.97$ according to the fit model in eq.~\eqref{eq:threeptfitfunc}.}
    \label{fig:fit_43}
\end{figure}

We report the results of the fits in tab.~\ref{tab:fit_res_baryon} and, as an example, we show in fig.~\ref{fig:fit_43} the best fit at $\beta=42.97$.

\begin{table}[]
\centering
\resizebox{\textwidth}{!}{%
\begin{tabular}{|l|l|l|l|l|l|l||l|l|}
\hline
$\beta$ & $R_{min}/a$ & $R_{max}/a$ & $A_3\times10^{6}$          & $\sigma \, a^2$ & $aM$         & $\chi^2/N_{d.o.f}$ & $\sigma_0a^2$ from ref.~\cite{Caselle:2024zoh} &$M/\sqrt{\sigma}$ \\ \hline
$36.33$ & 4.64        & 12.68       & 7.85(58) & 0.009022(53) & 0.0137(13)  & 1.19  &  0.009344(17)  & 0.142(13)  \\ \hline
$39.65$ & 4.64        & 13.93       & 4.75(18) & 0.007518(23) & 0.01210(53) & 1.38  &  0.007831(58)  & 0.137(6)   \\ \hline
$42.97$ & 4.64        & 13.93       & 3.11(11) & 0.006378(18) & 0.01052(38) & 0.98  &  0.006645(59)  & 0.129(5)   \\ \hline
$46.29$ & 5.87        & 15.17       & 2.00(14) & 0.005460(27) & 0.01007(91) & 0.97  &  0.005716(59)  & 0.133(12)  \\ \hline
\end{tabular}
}  
\caption{Result of the best fit of the Polyakov-loop three-point correlator $G^{(3)}(R)$ for different values of $\beta$, using the definition of $R$ of eq.~\eqref{eqRY}.}\label{tab:fit_res_baryon}
\end{table}

\begin{figure}[!htb]
    \centering
    \includegraphics[width=\linewidth]{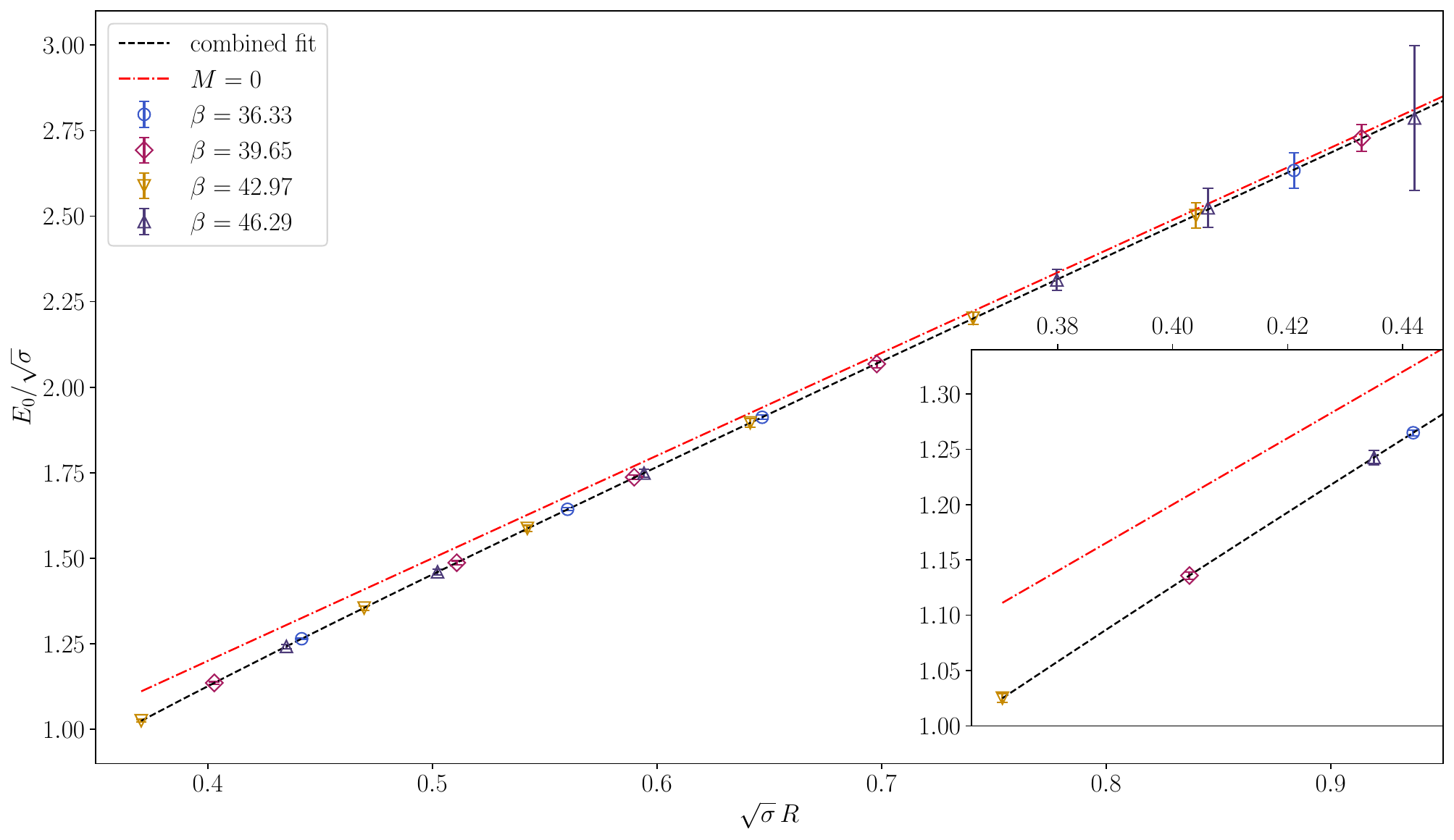}
    \caption{Combined best fits of the data across all four values of $\beta$ according to the fit model in eq.~\eqref{eq:threeptfitfunc}, performed under the assumption of a fixed same value of $M / \sqrt{\sigma}$. The correlator data points have been converted into the ground state energy level according to eq.~\eqref{eqlowT}, $E_0 = -\log \left(\braket{P(x_1) \, P(x_2) \, P(x_3)} / A_3 \right) / N_t$. This quantity does not show any correction due to the finite lattice spacing. The values of $A_3$ and $\sigma$, have been extracted from the combined fit (as independent parameters for each value of $\beta$). The red dash-dotted line represents the leading order expression for the ground state, assuming $M = 0$, \textit{i.e.} $E_0 = 3 \sigma \, R$. The zoomed inset shows the points at the smallest distances, it is evident how the inclusion of the term proportional to $M$ is indispensable for an accurate description of the data.} 
    \label{fig:global_fit}
\end{figure}

Since the ratio $M/\sqrt{\sigma}$ exhibits no significant variation with $\beta$, we performed a global fit across all four $\beta$ values, fixing a single common value of $M/\sqrt{\sigma}$ for every data set. We treated the string tensions $\sigma \, a^2$ for each $\beta$ as the only remaining free parameters.  This combined fit reduces the total number of independent parameters and yields to a more precise and stable determination of $M/\sqrt{\sigma}$. The global fit is shown in fig.~\ref{fig:global_fit} and the result, which represents our final estimate of the baryon junction mass, is the following:
\begin{equation}
    \frac{M}{\sqrt{\sigma}}= 0.1355(36) ,\;\;\;\; \chi^2/N_{d.o.f}=1.04.
\end{equation}
Besides its phenomenological interest this value, and in particular its sign, can be used to test existing models based on the AdS/QCD duality for a non-perturbative description of confinement (for examples of this class of models see for instance \cite{Andreev:2015riv,Andreev:2021bfg,Andreev:2025ekv}).
 
\subsubsection{String tension determination}
A rather non-trivial prediction of the EST model is that the string tension extracted from the 
two-point function (the $\braket{P^+(0)P(R)}$ correlator) should coincide with that extracted form the Baryonic state.
As shown in tab.~\ref{tab:fit_res_baryon}, we found a small, but significant deviation between the two values in our data. 
The string tension $\sigma$ obtained from the best fit is systematically lower by about $3–4\%$ with respect to the value extracted from the  $\braket{P^+(0)P(R)}$ correlator (reported in the eighth column in tab.~\ref{tab:fit_res_baryon}). It is interesting to notice that a similar discepancy has been observed recently in other high precision simulations of Baryonic states in (3+1) dimensions \cite{Ma:2022vqf}. Although this discrepancy is relatively small, it is however,  significant in view of the high precision of our data. As we shall see below, a similar discrepancy is also present in our high temperature results.

\subsection{Analysis of the high $T$ data}
\label{results_SY}
We used our high temperature results to test the two issues discussed at the end of sect.~\ref{comparison}, suggested by the Svetitsky--Yaffe mapping 

To this end, we performed independent fits with the large distance (eq.~\eqref{eq12}) and short distance (eq.~\eqref{eqsd}) expansions suggested by the mapping. Details on the data that we used are reported in  tab.~\ref{tab:simdetSY}.

More precisely, we fitted the large distance regime with:
\begin{equation}\label{eq:long-distance}
G^{(3)}(R_Y) = A_3 \, K_0(R_Y E_0),
\end{equation} 
using $A_3$ and $E_0$ as free parameters.

In the short-distance regime, we truncated the expression of eq.~\eqref{eqsd} to the order $r^{6/5}$ since the next term was compatible with zero within the precision of our data.
\begin{equation}\label{eq:low-distance} 
G^{(3)}(r) = \frac{A_0}{r^{2/5}} \left( c_1 + c_2 r^{4/5} + b r^{6/5} \right),
\end{equation} 
where $r = E_0 l$ and $l = \frac{R_Y}{\sqrt{3}}$ is the length of the side of the equilateral triangle. In this case, $A_0$, $E_0$, and $b$ are the free parameters, while $c_1$ and $c_2$ are fixed constants from \cite{Caselle:2005sf}, with values $c_1 = C_{\sigma \sigma}^{\bar{\sigma}} = 1.09236\dots$ and $c_2 = C_{\sigma \sigma \sigma}^{\epsilon} A_{\epsilon} \kappa^{2/3} = -2.2386\dots$. The coefficient $A_\epsilon$ was determined from the short-distance analysis of the two-point Polyakov loop correlator reported in \cite{Caselle:2024zoh}.

This ansatz for our fit corresponds to assuming only the universal terms and ratios to be the same as in the Potts model. In particular, $b$ is left as a free parameter since we cannot exclude that the Yang--Mills theory includes operators contributing to the $r^{6/5}$ terms with a different VEV from those present in the Potts model. Also the $c_2$ coefficient is, in principle, not universal by itself either, but its ratio with the energy VEV (estimated from the two point function in \cite{Caselle:2024zoh}) is so.

\begin{figure}[!htb]
    \centering
    \includegraphics[width=\linewidth]{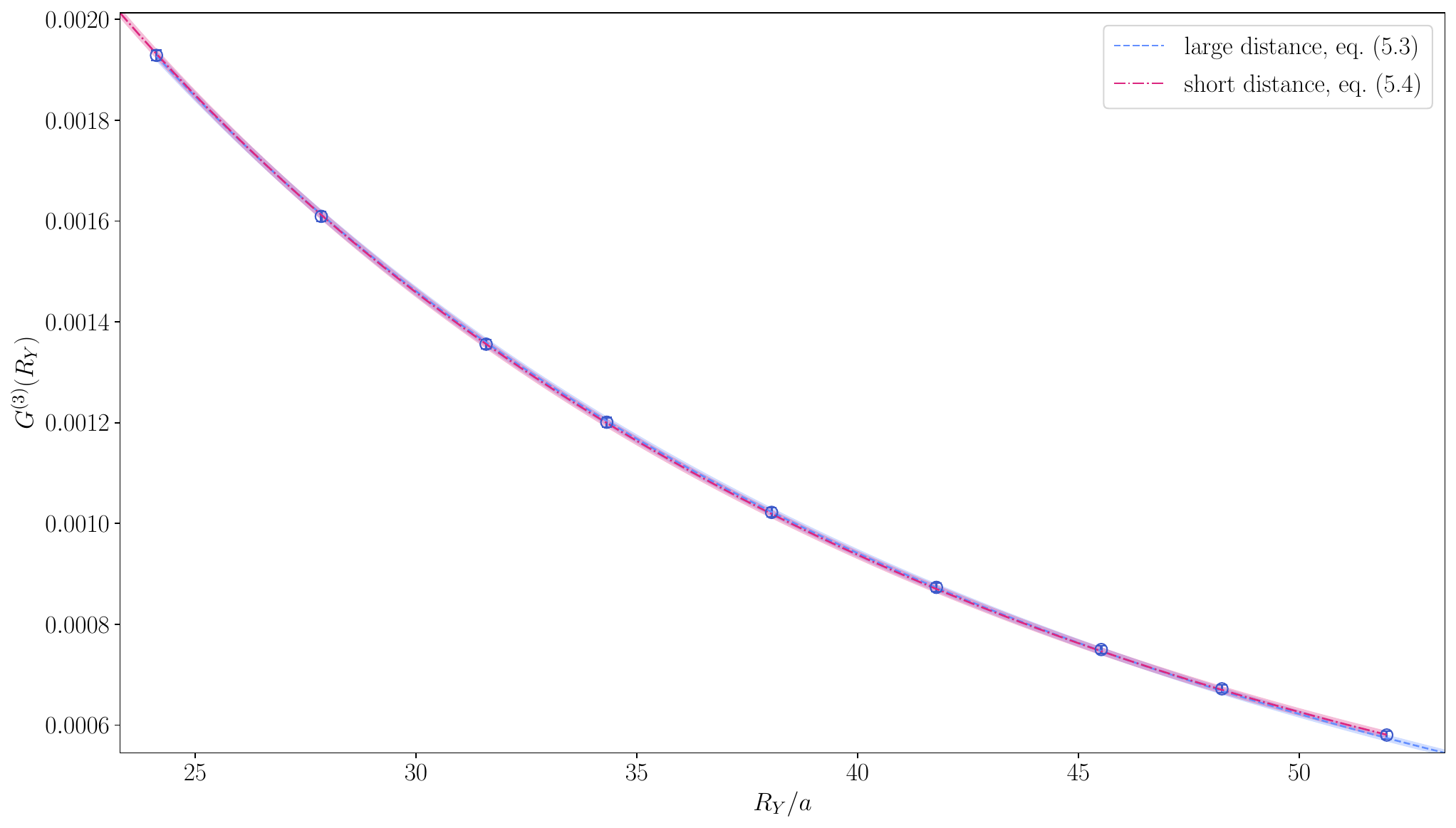}
    \caption{Fit of the three-point correlation function for $\beta = 46.29$ and $N_t = 15 \, a$ with the two models in eq.~\eqref{eq:low-distance} (short distance expansion, blue dashed line in the plot) and eq.~\eqref{eq:long-distance} (large distance expansion, red dashed-dotted). In this region ($24.12 \le R_Y / a \le 51.96$), they both fit accurately our data (blue circles) and show perfect agreement between each other.}
    \label{fig:overlap_log_short}
\end{figure}

In fig.~\ref{fig:overlap_log_short}, we show the remarkable agreement of the long and short distance expansion in the region where they both apply, at the order we considered them, for $\beta = 46.29$ and $N_t = 15 \, a$ as an example.

\begin{table}[!htb]
\centering
\begin{tabular}{|l|l|l|l|l|l|l|l|l|}
\hline
$\beta$                  & $N_t / a$               &  & $R_Y^{\text{(min)}} / a$ & $R_Y^{\text{(max)}}/a$ & Amplitude     & $aE_0$      & $b$        & $\chi^2/N_{d.o.f}$ \\ \hline
\multirow{2}{*}{$36.33$} & \multirow{2}{*}{12} & \eqref{eq:low-distance}  & 20.39           & 38.05           & 0.003477(51)  & 0.04410(55) & 1.2916(22) & 0.91               \\ \cline{3-9} 
                         &                     & \eqref{eq:long-distance}  & 27.86           & 69.64           & 0.003330(52) & 0.04504(44) &            & 1.4               \\ \hline
\multirow{4}{*}{$39.65$} & \multirow{2}{*}{13} &  \eqref{eq:low-distance} & 20.39           & 41.78           & 0.003375(43)  & 0.03864(46) & 1.2927(25) & 0.95               \\ \cline{3-9} 
                         &                     &  \eqref{eq:long-distance} & 24.12          & 100.23          & 0.003389(33) & 0.04035(32) &            & 1.04             \\ \cline{2-9} 
                         & \multirow{2}{*}{14} & \eqref{eq:low-distance}  & 20.39           & 34.32           & 0.002958(45)  & 0.05116(56) & 1.2804(16) & 0.73               \\ \cline{3-9} 
                         &                     &  \eqref{eq:long-distance} & 24.12         & 100.23          & 0.003167(33) & 0.05671(32)&            & 0.80               \\ \hline
\multirow{6}{*}{$42.97$} & \multirow{2}{*}{14} &  \eqref{eq:low-distance} & 20.39           & 51.98           & 0.003220(36)  & 0.03328(38) & 1.2950(26) & 1.36               \\ \cline{3-9} 
                         &                     & \eqref{eq:long-distance}  & 27.86           & 100.23          & 0.003144(43) & 0.03404(39) &            & 1.3               \\ \cline{2-9} 
                         & \multirow{2}{*}{15} &  \eqref{eq:low-distance}  & 20.39           & 34.32           & 0.002978(54)  & 0.04628(69) & 1.2883(27) & 0.79               \\ \cline{3-9} 
                         &                     &  \eqref{eq:long-distance}  & 20.39          & 100.23          & 0.003098(23) &0.04971(27) &            & 0.74               \\ \cline{2-9} 
                         & \multirow{2}{*}{16} &  \eqref{eq:low-distance} & 24.12           & 38.05           & 0.002305(47)  & 0.05052(62) & 1.2691(14) & 1.3                \\ \cline{3-9} 
                         &                     &  \eqref{eq:long-distance} & 20.39          & 100.23          & 0.002900(22) &0.06192(27)  &            & 0.77               \\ \hline
\multirow{8}{*}{$46.29$} & \multirow{2}{*}{15} & \eqref{eq:low-distance}  & 20.39           & 51.98          & 0.003167(33)  & 0.03023(36) & 1.2981(31) & 0.77               \\ \cline{3-9} 
                         &                     &  \eqref{eq:long-distance} & 24.12       & 86.30           & 0.003205(28) & 0.03150(29)&            & 0.9               \\ \cline{2-9} 
                         & \multirow{2}{*}{16} & \eqref{eq:low-distance}  & 20.39           & 38.05           & 0.002942(38)  & 0.04178(48) & 1.2906(23) & 0.63               \\ \cline{3-9} 
                         &                     &  \eqref{eq:long-distance} & 20.39           & 100.23          & 0.003051(19) &  0.04464(22) &            & 0.60               \\ \cline{2-9} 
                         & \multirow{2}{*}{17} & \eqref{eq:low-distance} & 24.12           & 38.05           & 0.002394(52)  & 0.04670(68) & 1.2753(20)  & 1.5                \\ \cline{3-9} 
                         &                     &  \eqref{eq:long-distance} & 20.39           & 100.23          &0.002798(19) & 0.05459(24)  &            & 0.76              \\ \cline{2-9} 
                         & \multirow{2}{*}{18} &  \eqref{eq:low-distance} & 24.12           & 38.05           & 0.002071(40)  & 0.05195(57) & 1.2698(11) & 1.2                \\ \cline{3-9} 
                         &                     &  \eqref{eq:long-distance}  & 20.39           & 100.23          & 0.002592(19)& 0.06339(26)  &            & 0.75              \\ \hline
\end{tabular}
\caption{Results of the fits to the short- and long-distance behavior of the Polyakov-loop three-point correlator $G^{(3)}(R_Y)$, described by the model in eq.~\eqref{eq:low-distance} and eq.~\eqref{eq:long-distance}, respectively, following the Svetitsky–-Yaffe mapping, for various values of $\beta$ and $N_t$.}
\label{tab:highT}
\end{table}

Let us discuss a few important features of these results.

\begin{figure}[htb]
	\centering
	\includegraphics[width=\linewidth]{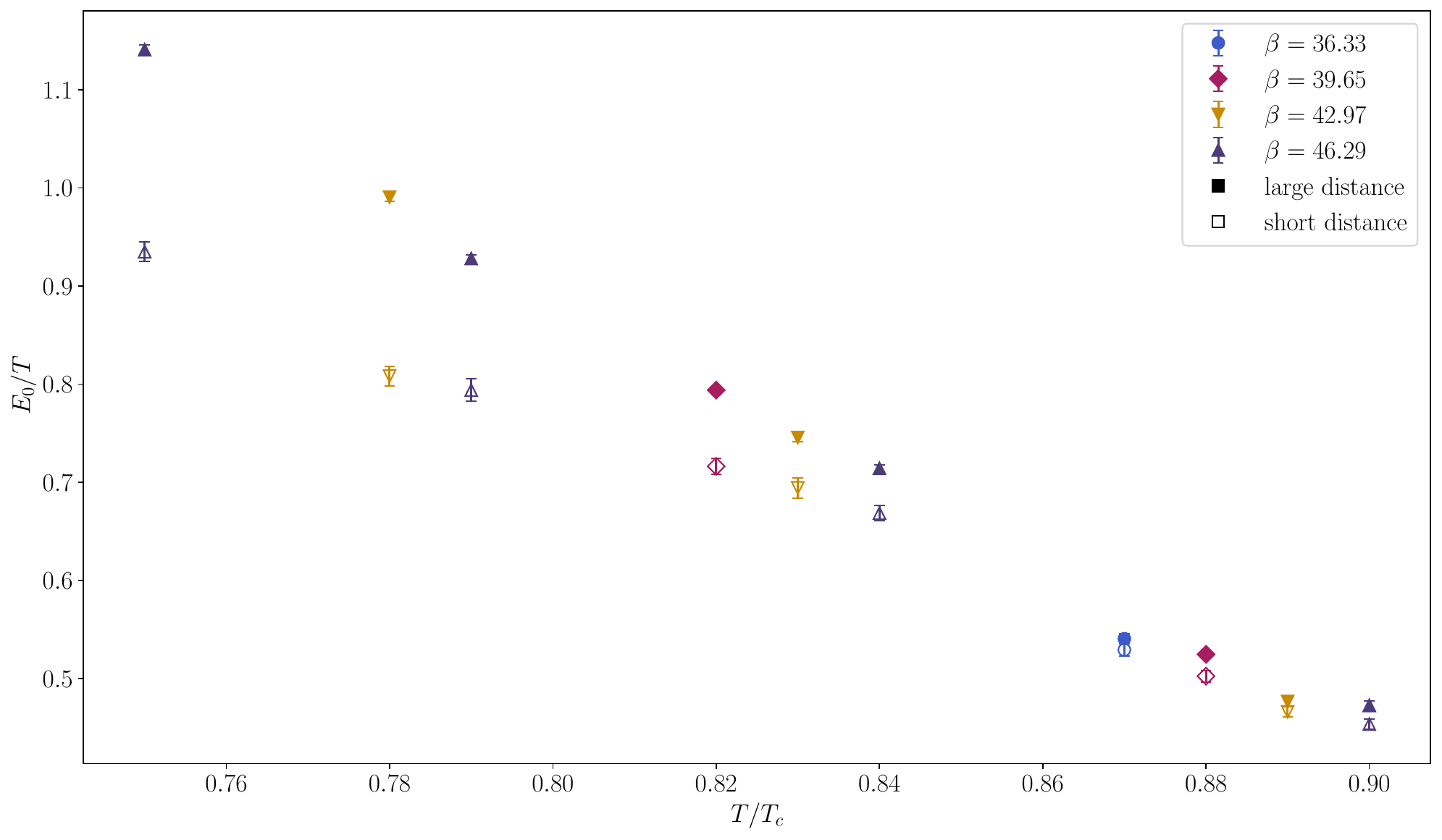}
	\caption{Ground state of the three-point correlation function at high temperature, extracted from fits assuming the short (empty bullets) and large (filled bullets) distance expansion at different values of $\beta$. The numerical values are reported in the seventh column of tab.~\ref{tab:highT}. It is evident how, approaching the critical temperature, the two estimates become compatible with each other.}
	\label{fig:ground_highT}
\end{figure}

A first observation is that the two values of $E_0$, extracted from the short and large distance expansions, get closer approaching the deconfinement temperature, as shown in fig.~\ref{fig:ground_highT}. The two estimations agree within the errors for the highest values of the temperature (above $T/T_c\simeq 0.85$) for all four values of $\beta$ that we studied. This means that for these values we have a prediction for the  functional form of the correlator of three Polyakov loops, which holds in the whole range of distances. In principle, we could use the functional form of the short-distance expansion as a tool to infer higher-order EST corrections, which become important only at very short distances. 

Another point is that the coefficient $b$, appearing in the short-distance fit, is approximately constant across the whole range of temperatures and lattice spacings that we studied and is compatible within the errors with the value $b=1.24(10)$  obtained in the analysis of the three-state Potts model, in ref.~\cite{Caselle:2005sf} for the same parameter , showing how small the deviations from the Potts model are, even in the non universal terms. A similar agreement was observed in ref.~\cite{Caselle:2024zoh} for the two-point function. This agreement represents a remarkable test of the Svetitsky--Yaffe mapping.

\begin{table}[!htb]
\centering
\begin{tabular}{|l|l|l|l|l|l|}
\hline
$\beta$ & $N_t / a$     & $\sigma a^2$ from eq.~\eqref{eq8}  & $\sigma a^2$ from eq.~\eqref{eq9}      & $\sigma a^2$ from tab.~\ref{tab:fit_res_baryon}& $\sigma a^2$ from ref.~\cite{Caselle:2024zoh}  \\ \hline
 36.33  & 12        & 0.007389(37)        & 0.008862(26)   & 0.009022(53) &   0.009344(17)      \\ \hline
\multirow{2}{*}{39.65}
        & 13        & 0.006202(25)        & 0.007484(17)   & 0.007518(23) &   0.007831(58)      \\ \cline{2-6}
        & 14        & 0.006722(23)        & 0.007524(19)   & 0.007518(23) &   0.007831(58)      \\ \hline
\multirow{3}{*}{42.97}
        & 14        & 0.005103(28)        & 0.006284(19)   & 0.006378(18) &   0.006645(59)      \\ \cline{2-6}
        & 15        & 0.005641(18)        & 0.006377(15)   & 0.006378(18) &   0.006645(59)      \\ \cline{2-6}
        & 16        & 0.005915(17)        & 0.006423(15)   & 0.006378(18) &   0.006645(59)      \\ \hline
\multirow{4}{*}{46.29}
        & 15        & 0.004427(19)        & 0.005462(13)   & 0.005460(27) &   0.005716(59)      \\ \cline{2-6}
        & 16        & 0.004835(14)        & 0.005505(11)   & 0.005460(27) &   0.005716(59)      \\ \cline{2-6}
        & 17        & 0.005023(14)        & 0.005499(12)   & 0.005460(27) &   0.005716(59)      \\ \cline{2-6}
        & 18        & 0.005138(14)        & 0.005491(13)   & 0.005460(27) &   0.005716(59)      \\ \hline
\end{tabular}
\caption{Values of $\sigma a^2$ extracted from the results for $E_0$ obtained using the large distance fits reported in tab.~\ref{tab:highT}.\label{tab:sigma}. In the third column, we report the results obtained assuming for $E_0$ the expression of eq.~\eqref{eq8}: $E_0(N_t)=\sigma N_t\left(1-\frac{\pi}{6\sigma N_t^2}\right)$.  In the fourth column, those obtained with the expression of eq.~\eqref{eq9}: $E_0(N_t)=\sigma N_t\sqrt{1-\frac{\pi}{3\sigma N_t^2}}$. In the last two columns, we quote as a reference the values obtained at low temperature (see tab.~\ref{tab:fit_res_baryon}) and those taken from ref.~\cite{Caselle:2024zoh}, looking at the ground state energy of the two-point function.}
\label{tab5}
\end{table}

Moreover,  from the values of $E_0$ of the large-distance fits we may extract the string tension. We report in the third column of tab.~\ref{tab5} the values obtained assuming for $E_0$  the result of the first order EST calculation of \cite{Komargodski:2024swh}: $E_0(N_t)=\sigma N_t\left(1-\frac{\pi}{6\sigma N_t^2}\right)$. In the same table, in the last two columns, we report for comparison the values obtained in the low temperature fits discussed above (fifth column) and those taken from the literature (last column). It is easy to see that the values of the third column disagree with the reference ones and that the disagreement increases as the deconfinement transition is approached. This is not surprising, it is due to the fact that the EST calculation of \cite{Komargodski:2024swh} is only a first-order calculation, and indeed the same happens also for the two-point correlator in this regime \cite{Caselle:2024zoh}. The Svetitsky--Yaffe mapping suggests that the exact EST result, resummed to all orders, should be  $E_0(N_t)=\sigma N_t\sqrt{1-\frac{\pi}{3\sigma N_t^2}}$ (see the discussion in sect.~\ref{comparison} above). The values obtained in this way for $\sigma a^2$ are reported in the fourth column of tab.~\ref{tab5} and show in general a good agreement with the reference values of the fifth column.

\begin{figure}[htb]
    \centering
    \includegraphics[width=\linewidth]{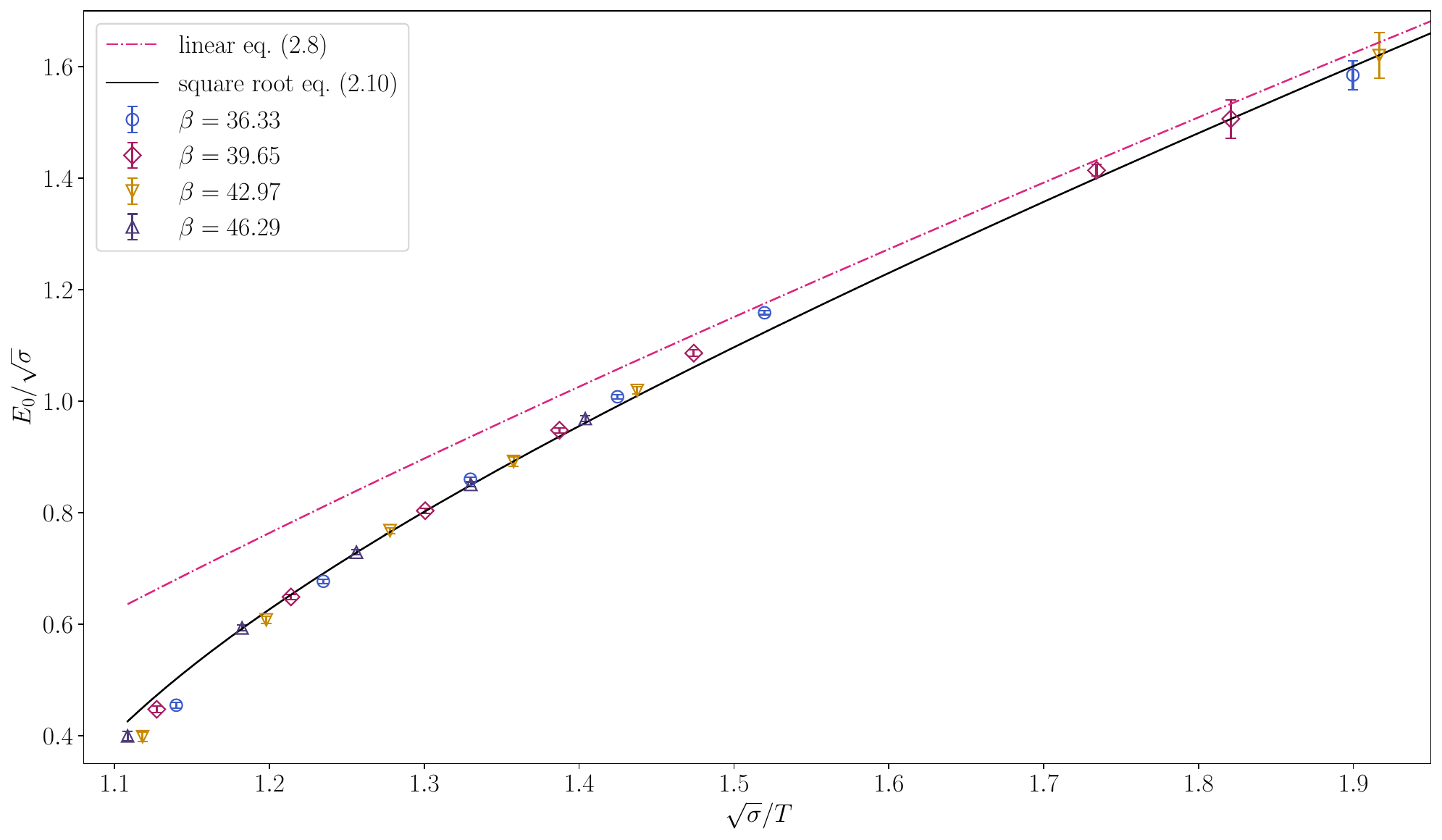}
    \caption{The value of $E_0$ obtained at high temperature from the fit with the large distance expansion in eq.~\eqref{eq:long-distance}. In order to set the scale on both axes, we used the value of the string tension obtained at the same lattice spacings as the data we are plotting at low temperature. In these units, both the proposed model from eq.~\eqref{eq8} and eq.~\eqref{eq9} have no free parameters, thus we plotted them as a dash-dotted red line and a dashed black line, respectively. The data show good agreement with the latter, suggesting that the string tension has the same value in the closed (corresponding to high temperature) as in the open channel (probed at low temperature) and that the square root ansatz is a better approximation of the true ground state.}
    \label{fig:high_T_scaled}
\end{figure}

In order to prove that this intuition is well founded, we plot the values of $E_0$ obtained at high temperature, scaled by the square root of the values of $\sigma$ obtained at low temperature. As we can see in fig.~\ref{fig:high_T_scaled}, they are in good agreement with the square root formula in eq.~\eqref{eq9}, while the linear ansatz of eq.~\eqref{eq8} shows significant deviations close to the phase transition. We also performed a fit according to the model of eq.~\eqref{eq9}, whose results are reported in Appendix \ref{App:D}, together with some additioonal data.

Furthermore, we have seen in the section on the low-T fits that there is a slight discrepancy of the values of $\sigma a^2$ obtained from the three-point function (fifth column of tab.~\ref{tab5}) and those obtained from the two-point function (last column of tab.~\ref{tab5}) which are systematically larger.
It is interesting to see that the values that we obtain in the high temperature fits agree well with those obtained in the low temperature fits and not with those of ref.~\cite{Caselle:2024zoh}. It is, in principle, possible that this discrepancy at high temperature is due to  a difference  in the corrections to the Nambu--Got\=o closed string expression of eq.~\eqref{eq8bis} between the two- and three-point functions.  However, this would explain the discrepancy only in the high temperature regime, while we observe the same discrepancy also in the low temperature regime. This slight discrepancy seems thus to be a structural feature of the string tension extracted from the three-point function.  It also agrees with some recent independent observation performed in 3+1 dimensions \cite{Ma:2022vqf}.  It would be interesting to understand if this effect is related in some way to the presence of the baryon junction, for instance, as a consequence of the  broadening of the flux tube in the vicinity of the baryon junction discussed in ref. \cite{Pfeuffer:2008mz}. We plan to further address this point in future simulations.

Finally, looking at the tab.~\ref{tab:highT} we see that the large distance fits converge down to rather small values of $R_Y$. In general, one would expect the large distance approximation to be reliable only for values larger than the correlation length $1/E_0$, which is typically a much larger value. We think that the reason behind this unexpected good behavior at short distance is exactly the fact that the large distance expansion coincides with the EST approximation, whose critical radius $\sim\frac{1}{\sqrt\sigma}$, which is in fact much smaller than $1/E_0$.

\section{Conclusions}
\label{CONCLUSION}

In this work, we carried out a systematic study on the lattice of a three–string baryonic configuration in three-dimensional $\SU(3)$ Yang--Mills theory. 
By exploiting the next-to-leading-order expressions derived in ref.~\cite{Komargodski:2024swh}, within the Effective String Theory (EST) framework, and performing high precision lattice simulations for the three-point Polyakov–loop correlator, we were able to determine for the first time the Baryon Junction mass $M$. This quantity has important consequences both on the theoretical side and on the phenomenological side. 
Our result is:
\begin{equation}
    \frac{M}{\sqrt{\sigma}}=0.1355(36).
\end{equation}
This estimate turns out to be similar to the phenomenological value which is used to describe hadrons (see, for instance, \cite{Karliner:2016zzc}). Needless to say that this agreement is, at this stage, nothing more than a suggestive coincidence, since we are dealing with the $2+1$ dimensional model. To test if the agreement persists also in $3+1$ dimensions it is mandatory to repeat the computation directly in the $(3+1)$ dimensional theory. This is a much more complex task than the one we studied in this paper and we plan to address it in a forthcoming publication.

The numerical value that we obtained for the baryon junction mass and, in particular, the sign of this quantity also indicate that the EST describing this configuration lies in the weakly coupled, perturbatively stable and in an unitary regime~\cite{Komargodski:2024swh}, and can be used to test existing models based on the AdS/QCD duality for a non-perturbative description of confinement.

A second major outcome of our study concerns the high-temperature regime, just below the deconfinement point.  We compared the data for the baryonic correlator with the quantitative predictions obtained using conformal perturbation theory in the three-state Potts model in one dimension less. The lattice data display excellent agreement with this independent description, based on the Svetitsky–-Yaffe mapping. This complements and extends to the three-point function the similar results recently obtained for the two-point function, discussed in \cite{Caselle:2024zoh}. 

It would be interesting to study the  broadening of the flux tube in the vicinity of the baryon junction and test the predictions of ref. \cite{Pfeuffer:2008mz}. This could be done, for instance applying the methods discussed in \cite{Verzichelli:2025cqc} and would represent an independent, stringent test of the validity of the EST description of Baryons.

Moreover, a natural follow-up is to extend our calculation to the $\SU(3)$ Yang-Mills theory in (3+1) dimensions and to full QCD to test if the   $\frac{M}{\sqrt{\sigma}}$ remains of the same order of magnitude. In this respect, it is interesting to notice that it has been recently shown in a simplified model that several features of the EST survive the coupling with matter \cite{Bonati:2021vbc}.

\vskip 1.5cm
\noindent {\large {\bf Acknowledgments}}
\vskip 0.2cm
We thank A. Bulgarelli, T. Canneti, E. Cellini, A. Mariani, A. Nada, and M. Panero for several useful discussions, and Z. Komargodski and O. Andreev for insightful discussions and a careful reading of the first version of the draft.  

We acknowledge support from the SFT Scientific Initiative of INFN. This work was partially supported by the Simons Foundation grant 994300 (Simons Collaboration on Confinement and QCD Strings) and by the Prin 2022 grant 2022ZTPK4E. The numerical simulations have been performed on the Leonardo machine at CINECA, based on the agreement with INFN, under the project INF25\_sft.
\vskip 1cm

\appendix

\section{Details on the three spin correlator in the three-states Potts model.}
\label{App:SY}
\subsection{Short distance expansion}

For all the details we address the interested reader to the original literature    
\cite{Caselle:2005sf,Guida:1995kc}.   

The key ingredient of the Conformal perturbation approach of \cite{Guida:1995kc} is to express the perturbed correlation functions in terms of integrals on the correlation functions evaluated at the conformal point and Vacuum Expectation Values $\langle [\phi_p] \rangle$ of the operators of the theory involved in the expansions   
These VEVs cannot be calculated in the framework of the perturbation theory, and being non-perturbative objects they have to be obtained by other methods. In our case, thanks to the exact integrability of the three-states Potts model these VEVs can be computed analytically. They were computed in a series of papers \cite{Lukyanov:1996jj, Fateev:1997yg, Fateev:1998xb, Fateev:1993av, Zamolodchikov:1989cf} (not only for our model, but for a wide class of theories, including various  integrable perturbations of the Minimal Models).   
   
Even if we did not use the two-point function in this paper, let us start our analysis by studying this case, which is simpler and allows to understand the logic of the expansion.

\subsection{Two-point function}
The leading term of the perturbative expansion is given by the two-point conformal correlator:    
\bea   
{\mathcal C}_{\sigma {\bar \sigma}}^{ \mathbb I}(x;0)= \frac{1}{|x|^{4/15}}   
\label{2pt} .
\eea   

For the next to leading terms we need the VEV of the perturbing operator $\epsilon(x)$ and the relation between the coupling constant and the mass of the fundamental particle. The latter is known, and can be extracted from \cite{Fateev:1993av}   
\bea   
\tau = \kappa\, m^{6/5}, \ \ \ \ \kappa = 0.164303 \dots.   
\label{kappa} 
\eea   
The VEV $\langle \epsilon \rangle$ can be computed starting from the knowledge of the vacuum energy density \cite{Zamolodchikov:1989cf}   and turns out to be  
\bea    
\langle \epsilon \rangle = - 2.92827\dots m^{4/5}.   
\eea   
Collecting the above ingredients, the perturbative series can be cast in the following form   
\bea   
\langle \sigma(0) {\overline{\sigma}}(R) \rangle  =   \frac{A_\sigma^2}{|r|^{4/15}} \left( 1 + g_1 r^{4/5} + g_2 r^{6/5}  \cdots \right), 
\eea    
  
 where $r=m R$ and the two constants $g_1=-1.59936\dots$ and $g_2 = 0.805622 \dots$ are obtained by a suitable combination of VEVs, OPE coefficients and integrals of conformal correlators  (see \cite{Caselle:2005sf} for details) and $A_\sigma$ is a non-universal normalization.
 
This expansion was used in a recent paper \cite{Caselle:2024zoh} to fit the short distance behavior of mesonic potential in the SU(3) gauge model in (2+1) dimensions in a way similar to what we are doing here for the baryonic potential.

\subsection{Extension to multi-point functions: general framework}     
Our main interest in this paper is the conformal perturbation of the three-point function which we use to describe the short distance behavior of the Baryon potential at high temperature. The remarkable feature of the perturbative approach proposed in \cite{Guida:1996nm} is that it can be straightforwardly generalized to multi-point correlators. Following the guidelines of \cite{Guida:1996nm} one can write a perturbative expansion for the multi-point function as follows:  
\bea   
G^{(n)} (x) =  \langle \phi_1(x_1) \dots \phi_n(x_n) \rangle =    
\sum_p {\mathcal C}_{ \phi_1 \dots \phi_n}^{[\phi_p]}    
(\underline{x}; \tau)    
\langle [\phi_p] \rangle   ,
\eea   
where the structure functions ${\mathcal C}_{ \phi_1 \dots \phi_n}^{[\phi_p]} (\underline{x}; \tau) $ are a generalization of those appearing in the formula for the two-point function\footnote{We used the notation $\underline{x} = \{x_1, \dots, x_n\}$.}. Their Taylor expansion gives   
\bea   
{\mathcal C}_{ \phi_1 \dots \phi_n}^{[\phi_p]} (\underline{x}; \tau) =    
{\mathcal C}_{ \phi_1 \dots \phi_n}^{[\phi_p]} (\underline{x}; 0) +   
\sum_{\ell =1}^{\infty} \frac{\tau^\ell}{\ell !}    
\partial_\tau^\ell {\mathcal C}_{ \phi_1 \dots \phi_n}^{[\phi_p]} (\underline{x}; 0)    
\eea   
where we have   
\bea   
{\mathcal C}_{ \phi_1 \dots \phi_n}^{[\phi_p]} (\underline{x}; 0) =    
\langle \phi_1(x_1) \dots \phi_n(x_n) [\phi_p](\infty)  \rangle_{\textrm{\tiny cft}},   
\eea   
and   
\bea   
\partial_\tau^\ell {\mathcal C}_{ \phi_1 \dots \phi_n}^{[\phi_p]} (\underline{x}; 0)=(-1)^{\ell} 
\int^{'} d^2 z_1 \dots d^2 z_\ell \,    
\langle \phi_1(x_1) \dots \phi_n(x_n) \epsilon(z_1) \dots \epsilon(z_\ell)   
 [\phi_p](\infty)  \rangle_{\textrm{\tiny cft}}   
\eea   
where the operator $\epsilon(x)$ which appears in the previous expression is the perturbing operator conjugated to the coupling constant $\tau$.

\subsection{Three-point function for the three-states Potts model}     
We are interested in particular in the 3-point function $G^{(3)} = \langle \sigma_1 \sigma_2 \sigma_3 \rangle$, for the three-states Potts model in two dimensions.  As for the 2-point function above, we can leverage in this case several exact results due to the exact integrability of the model. The perturbative expansion of  $G^{(3)}$ can be written according to the previous considerations as    
\bea   
G^{(3)} (x) &  = & \langle \sigma(x_1) \sigma(x_2)  \sigma(x_3) \rangle =   
\nonumber \\   
& = &    
{\mathcal C}_{ \sigma \sigma \sigma}^{{\mathbb I}}    
(x_1,x_2,x_3; \tau) +{\mathcal C}_{ \sigma \sigma \sigma}^{\epsilon}    
(x_1,x_2,x_3; \tau) \, \langle \epsilon \rangle+    
\dots  \,\,; 
\eea    
up to first order in $\tau$ one has   
\bea   
{\mathcal C}_{ \sigma \sigma \sigma}^{{\mathbb I}}    
(x_1,x_2,x_3; \tau) & = &   
{\mathcal C}_{ \sigma \sigma \sigma}^{{\mathbb I}}    
(x_1,x_2,x_3; 0) +  \tau\, \partial_\tau {\mathcal C}_{ \sigma \sigma \sigma}^{{\mathbb I}}    
(x_1,x_2,x_3; 0) + \dots    
\nonumber \\   
{\mathcal C}_{ \sigma \sigma \sigma}^{\epsilon}    
(x_1,x_2,x_3; \tau) & = &   
{\mathcal C}_{ \sigma \sigma \sigma}^{\epsilon}    
(x_1,x_2,x_3; 0) +\tau\,  \partial_\tau   
{\mathcal C}_{ \sigma \sigma \sigma}^{\epsilon}     
(x_1,x_2,x_3; 0) + \dots   
\nonumber   
\eea   
The explicit expression of the zero-th order contributions can be derived quite easily. We have   
\bea   
{\mathcal C}_{ \sigma \sigma \sigma}^{{\mathbb I}}    
(x_1,x_2,x_3; 0) &=& \langle \sigma(x_1) \sigma(x_2)  \sigma(x_3)  \rangle_{\textrm{\tiny cft}} =    
\nonumber \\   
&=& \frac{C^{\bar \sigma} _{\sigma \sigma}}{|x_1-x_2|^{2/15}|x_2 -x_3|^{2/15}   
|x_1-x_3|^{2/15}}   
\label{3pt} 
\eea   
where the structure constant $C^{\bar \sigma} _{\sigma \sigma}$ can be found in the literature \cite{Dotsenko:1985hi, Klassen:1991dz, McCabe:1995uq}   
\bea   
C^{\bar \sigma} _{\sigma \sigma}=    
\sqrt{\frac{3 \sin2 \pi/5 }{\pi \sin \pi/5}}    
\frac{\Gamma(5/6)\Gamma^2(3/5)\Gamma^4(1/3)}{\Gamma(2/5)\Gamma(4/5)\Gamma^2(2/3)\Gamma^2(1/6)}   
=1.09236 \dots.   
\eea   
The other zero-th order term is given by   
\bea   
{\mathcal C}_{ \sigma \sigma \sigma}^{\epsilon}    
(x_1,x_2,x_3; 0)  =    
\langle \sigma(x_1)  \sigma(x_2) \sigma(x_3) \epsilon(\infty)  \rangle_{\textrm{\tiny cft}}   
\eea   
whose main ingredient is the conformal four-point correlator $\langle \sigma\sigma\sigma\epsilon \rangle$ which was computed in \cite{Gliozzi:1997yc}. We consistently fixed its normalization constant, which gives    
\bea   
\langle \sigma(x_1)  \sigma(x_2) \sigma(x_3) \epsilon(x_4)  \rangle_{\textrm{\tiny cft}}   
=C^{\bar \sigma} _{\sigma \sigma} C^{\epsilon} _{\sigma {\bar \sigma}} \,   
\frac{|x_{12}x_{13}x_{23}|^{2/15}}{|x_{14}x_{24}x_{34}|^{8/15}} \,   
|y(1-y)|^{14/15} \,   
\left \{   
|f_1(y)|^2 + K^{-1} |f_2(y)|^2   
\right \}   
\eea   
where   
\bea   
f_1(y) = y^{-3/5} {_2F_1}(1/5,4/5;2/5;y), \ \ \ f_2(y) = {_2F_1}(4/5,7/5;8/5;y)   
\eea   
and   
\bea   
K = \frac{9\, \Gamma(1/5) \, \Gamma(3/5)^3 }{4\, \Gamma(4/5) \, \Gamma(2/5)^3 }, \ \ \ y=\frac{x_{14}x_{23}}{x_{34} x_{21}}, \ \ \ 1-y=\frac{x_{13}x_{24}}{x_{34} x_{12}}   
\eea   
Then, a simple computation gives   
\bea   
{\mathcal C}_{ \sigma \sigma \sigma}^{\epsilon}    
(x_1,x_2,x_3; 0)  =    
C^{\bar \sigma} _{\sigma \sigma} C^{\epsilon} _{\sigma {\bar \sigma}} \,
|x_{12}x_{13}x_{23}|^{2/15} |{\tilde y}(1-{\tilde y})|^{14/15} \,   
\left \{   
|f_1({\tilde y})|^2 + K^{-1} |f_2({\tilde y})|^2   
\right \}   
\eea   
where   
\bea   
{\tilde y}=\frac{x_{23}}{x_{21}}, \ \ \ 1-{\tilde y}=\frac{x_{13}}{x_{12}}   
\eea   
The first order terms are as follows    
\bea   
\partial_\tau {\mathcal C}_{ \sigma \sigma \sigma}^{{\mathbb I}}    
(x_1,x_2,x_3; 0) & = & - \int^{'} d^2 z    
\langle \sigma(x_1)  \sigma(x_2) \sigma(x_3) \epsilon(z)  \rangle_{\textrm{\tiny cft}}   
\nonumber \\   
\partial_\tau {\mathcal C}_{ \sigma \sigma \sigma}^{{\epsilon}}    
(x_1,x_2,x_3; 0) & = &  - \int^{'} d^2 z   
\langle \sigma(x_1)  \sigma(x_2) \sigma(x_3) \epsilon(z)  \epsilon(\infty)  \rangle_{\textrm{\tiny cft}}   
\eea   
These expressions are quite complex for a generic triangle, but in the case of an equilateral triangle, which is the situation in which we are interested here, some relevant simplifications occur and allow an analytical calculation of the first few coefficients of the expansion.

\subsection{Three-point correlator: Equilateral triangles}      
Let $l$ be the side of the triangle. Then, we may choose (obviously the final results will be independent of such a choice) the relative position of the three points to be:
\bea   
x_1=0, \ \ \ x_2 = l \, e^{i \pi /6}, \ \ \ x_3 = l \, e^{-i \pi /6}   
\eea   
we have     
\bea   
{\tilde y}=i \, e^{-i \pi /6} , \ \ \ 1-{\tilde y}=e^{-i \pi /3}   
\eea   
and finally   
\bea   
{\mathcal C}_{ \sigma \sigma \sigma}^{\epsilon}    
(0, l \, e^{i \pi /6}, l \, e^{-i \pi /6}; 0)  =    
C^{\bar \sigma} _{\sigma \sigma} C^{\epsilon} _{\sigma {\bar \sigma}}    
 \,   
\left \{   
|f_1(i \, e^{-i \pi /6})|^2 + K^{-1} |f_2(i \, e^{-i \pi /6})|^2   
\right \}  l^{2/5} =  C^{\epsilon} _{\sigma \sigma \sigma } \, l^{2/5}    
\eea   
where   
\bea   
C^{\epsilon} _{\sigma \sigma \sigma } = 0.788825\dots.   
\eea   
Taking into account all the other pieces we have   
\bea   
G^{(3)}(l) =  \frac{1}{l^{2/5}} \left\{   
C^{\bar \sigma} _{\sigma \sigma}  +    
C^{\epsilon} _{\sigma \sigma \sigma } \, A_\epsilon \,   
\tau^{2/3} \, l^{4/5} + \dots   
\right\}.   
\eea   
This expression can be rewritten in terms of the dimensionless variable $r = m l$  
\bea   
G^{(3)}(l) =  \frac{1}{l^{2/5}} \left(   
c_1 + c_2 \, r^{4/5}   + \dots   
\right) 
\eea   
where 
\bea 
c_1 = C^{\bar\sigma} _{\sigma \sigma} = 1.09236\dots \;,\ \ \ \  
c_2 = C^{\epsilon} _{\sigma \sigma \sigma } \, A_\epsilon \, \kappa^{2/3} = -2.29795\dots 
\eea 

It is also useful to define the scale invariant form of the correlator  
\bea   
\widetilde{G}^{(3)}(r) =m^{-2/5}\, \langle \sigma(x_1)  \sigma(x_2)  \sigma(x_3)  
\rangle=  
\frac{1}{r^{2/5}} \left( 
c_1 + c_2 \, r^{4/5}  + \dots   
\right)  
\eea     
and its corresponding expression on the lattice 
\bea   
\widetilde{G}^{(3)}_{\text{lat}}(r) = 
\frac{A_{\sigma}^{3/2}}{r^{2/5}} \left( 
c_1 + c_2 \, r^{4/5}  + \dots   
\right) 
\eea 
where $A_{\sigma}$ is again  a non universal normalization constant.

A remarkable aspect of this perturbative approach is that the form of the higher order terms of expansion can be predicted relying on dimensional analysis only.
\bea   
\widetilde{G}^{(3)}_{\text{lat}}(r) = 
\frac{A_{\sigma}^{3/2}}{r^{2/5}} \left( 
c_1 + c_2 \, r^{4/5} + c_3 \, r^{6/5}  +c_4 \, r^2 + c_5 \, r^{12/5}  + \dots   
\right) 
\eea 
 
where $c_3, c_4, c_5, \dots$ are unknown (but in principle calculable) constants.  They were estimated numerically in \cite{Caselle:2005sf} to be: 
\begin{equation}
    c_3 = 1.24(10) \ , \ \   c_4 = 0.44(45) \ , \ \    c_5 = -0.33(37).
\end{equation}
This is the expression that we used in the main text to fit the short distance behavior of the baryon potential. As we have seen in the main text the last two terms are beyond the precision of our simulations and so we neglected them in our fits while $c_3$ could be determined rather precisely with our fits and turned out to agree with the estimate of \cite{Caselle:2005sf} remarkably well.

\section{Comparison of the different choices of the distance between the vertices and the Baryon junction on the lattice}
\label{App:B}
As mentioned in the paper, since the triangles are not exactly equilateral, the definition of $R$ in eq.~\eqref{eq:threeptfitfunc} is ambiguous. In particular, we need to clarify which point of the triangle is the "center", whose distance from the vertices is $R$. To be precise, $R$ appears twice in eq.~\eqref{eq:threeptfitfunc}; we will assume the corrections due to the non-equilateral triangles to be the same for the two instances. Possible differences between the corrections to the $R$ appearing in the term proportional to $M$ and that to the radius of the triangle are negligible, given the precision of our data. The picture of a "tight" string suggests to choose the center $O$, as the point that minimizes the sum of the distances of the vertices from it \cite{Bicudo:2008yr}. Thus, the value of $R$ will be one-third of the sum, denoted by $R_Y$, of the three distances from $O$ to the vertices (which are not exactly equal to each other).

It is known that for triangles with all the angles smaller than $2 \pi / 3$, such a point has the property that the segment which connect it to the vertices intersect forming three equal angles (of $2 \pi / 3$). Thus, for isosceles triangles, $O$ is easy to identify, since $O$, together with its reflection with respect to the base of the triangle and one endpoint of the base, will form an equilateral triangle. From this observation follows that the distance of $O$ from the base is $b / (2 \, \sqrt{3})$ and $b / \sqrt{3}$ from its endpoints, where $b$ is the length of the base. Hence, we obtain the formula:
\begin{equation}
    R_Y = \frac{b}{\sqrt{3}} + \frac{b}{\sqrt{3}} + \left( h - \frac{b}{2 \, \sqrt{3}} \right)
        = \frac{\sqrt{3}}{2} \, b + h,
\end{equation}
where $h$ is the height of the triangle. And
\begin{equation}
    R = \frac{R_Y}{3} = \frac{b}{2 \, \sqrt{3}} + \frac{h}{3},
\end{equation}
which of course becomes $b / \sqrt{3}$ if $h$ is exactly $\sqrt{3} / 2 \, b$.

Interesting alternative choices are the value of $R$ for an equilateral triangle of base $b$, which is $b / \sqrt{3}$, and the radius of the circumscribed circumference of the triangle, which is $b^2 / (8 \, h) + h / 2$.

To assess which definition of $R$ is most appropriate, we repeated the fits using different prescriptions and compared the quality of the results. In particular, we analyzed the normalized fit residuals (\textit{i.e.} the numerical data minus the value of the fitted function, divided by the statistical uncertainty) as a function of the \textit{equilateral deviation parameter}, which quantifies how much the triangle deviates from an equilateral shape. It is defined as  
\begin{equation}
    \label{eq:nonequilat}
    \delta_{\text{neq}} = \frac{b - l}{b},
\end{equation}
where $b$ is the length of the base of the triangle and $l$ is the length of the other two (equal) sides. This parameter vanishes for an equilateral triangle and increases with asymmetry.

\begin{figure}[!htb]
    \centering
    \includegraphics[width=\linewidth]{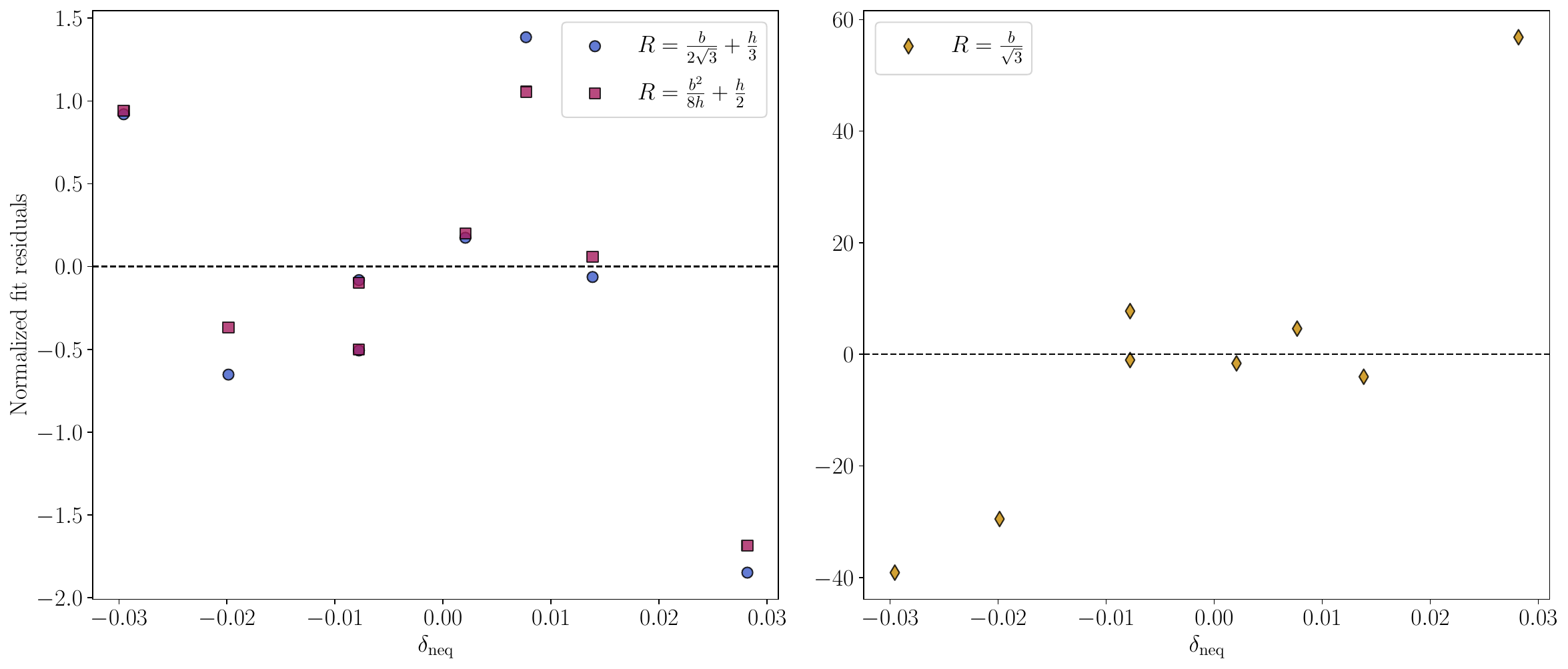}
    \caption{Normalized residuals of the best fits of the three-point correlation function $G^{(3)}(R)$ obtained using the three different definitions of the distance $R$, for the simulations at $\beta = 39.33$. In the left panel we show the residuals obtained using the definitions based on $R_Y$ (blue circle) and the one based on the radius of the circumscribed circumference of the triangle (red square), in the right panel we show the results adopting the naive definition valid for an equilateral triangle (note the different scale of the y axis). The horizontal axis shows the equilateral deviation parameter $\delta_{\text{neq}}$, as defined in eq.~\eqref{eq:nonequilat}, which quantifies the deviation from an equilateral triangle.}

    \label{fig:residuals_vs_neq}
\end{figure}

For instance, in fig.~\ref{fig:residuals_vs_neq}, we present the results of the residuals using the three different definitions of $R$ for the low temperature data at $\beta=39.33$. Clearly, the precision of our data is sufficient to evidentiate a positive correlation between $\delta_{\text{neq}}$ and the residuals from the fit function, when fitting with the $R = b / \sqrt{3}$ prescription.   

When adopting the naive definition $R = b/\sqrt{3}$ the residuals are generally large and exhibit a clear increasing trend as the triangle becomes less equilateral, i.e., as $ \delta_{\text{neq}}$ grows. On the other hand, using both the definitions based on $R_Y$, which accounts for the actual geometry of the triangle, and the one based on the radius of the circumscribed circumference of the triangle, the residuals remain small and do not show a significant dependence on $\delta_{\text{neq}}$. This behavior indicates that the second and the third definition provide a more accurate and robust description of the data, particularly for non-equilateral configurations. Although the two definitions of $R$ lead to qualitatively consistent results in our analysis, we adopt the definition based on $R_Y$.

\section{The Dedekind and Eisenstein functions and their modular transformations}
\label{App:C}
The Dedekind function is defined as 

\begin{equation}
\begin{aligned}
\eta(q)\equiv q^{\frac{1}{24}}\prod_{n=1}^\infty(1-q^n)
    \end{aligned}
\end{equation}
where $q=e^{2\pi i \tau}$ and in our setting $\tau=i\frac{N_t}{2R}$. 
Its modular transformation is:

\begin{equation}
\eta(q)\,=\,\sqrt{-i\Tilde{\tau}}\,\eta(\Tilde{q})
    \end{equation}
where  $\Tilde{\tau}=-\frac{1}{\tau}$.

\vskip 0.3cm
\noindent
The Eisenstein series of order $2$ is defined as  
\begin{equation}
E_{2}(q)\equiv 1-24\sum_{n=1}^\infty\frac{nq^n}{1-q^n} \,\,\ .
\end{equation}
Its modular transformation is:
\begin{equation}
E_2(q)=-\frac{6i}{\pi }\Tilde{\tau}+\Tilde{\tau}^2E_2(\Tilde{q}).
    \end{equation}
    
\section{Fits of numerical results in the high temperature regime}    
\label{App:D}

We report in tab.~\ref{tab:highT_appendix} some additional data, for the ground state $E_0$ extracted from the three-point function at low distance. Since the correlation length $1 / E_0$ is very small in lattice units, in these cases we were not able to perform the short distance analysis. For this reason these data are not useful for our arguments in sect.~\ref{results_SY} and thus we did not include them in the main text. However, they turn out to be helpful in order to extract a value of the string tension from the closed channel, in order to compare it to the open channel one.

\begin{table}[!htb]
\centering
\begin{tabular}{|l|l|l|l|l|l|l|}
\hline
$\beta$                & $N_t / a$ & $R_Y^{\text{(min)}} / a$ & $R_Y^{\text{(max)}} / a$ & Amplitude    & $aE_0$      & $\chi^2/N_{d.o.f}$ \\ \hline
\multirow{6}{*}{36.33} & 13        & 24.12                    & 100.23                   & 0.003185(38) & 0.06431(37) & 1.08               \\ \cline{2-7} 
                       & 14        & 17.66                    & 100.23                   & 0.003142(22) & 0.08216(30) & 0.92               \\ \cline{2-7} 
                       & 15        & 17.66                    & 100.23                   & 0.002878(25) & 0.09573(37) & 0.81               \\ \cline{2-7} 
                       & 16        & 13.93                    & 100.23                   & 0.002743(16) & 0.11004(31) & 1.27               \\ \cline{2-7}
                       & 20        & 20.39                    & 69.64                    & 0.00165(10)  & 0.1505(25)  & 0.73               \\ \cline{2-7}
                       & 21        & 24.12                    & 69.64                    & 0.00179(39)  & 0.1678(78)  & 0.74               \\ \hline
\multirow{7}{*}{39.65} & 15        & 17.66                    & 100.23                   & 0.003010(19) & 0.07065(26) & 0.82               \\ \cline{2-7} 
                       & 16        & 17.66                    & 100.23                   & 0.002816(21) & 0.08276(31) & 0.84               \\ \cline{2-7} 
                       & 17        & 13.93                    & 100.23                   & 0.002685(19) & 0.09528(37) & 0.83               \\ \cline{2-7}
                       & 20        & 20.39                    & 69.64                    & 0.001915(44) & 0.12260(90) & 1.08               \\ \cline{2-7}
                       & 21        & 24.12                    & 69.64                    & 0.00168(15)  & 0.1306(30)  & 0.55               \\ \cline{2-7}
                       & 23        & 24.12                    & 69.64                    & 0.00131(15)  & 0.1465(41)  & 1.09               \\ \cline{2-7}
                       & 24        & 27.86                    & 69.64                    & 0.00116(41)  & 0.155(11)   & 0.86               \\ \hline
\multirow{2}{*}{42.97} & 17        & 17.66                    & 100.23                   & 0.002732(30) & 0.07246(45) & 0.73               \\ \cline{2-7} 
                       & 18        & 17.66                    & 100.23                   & 0.002527(20) & 0.08232(32) & 0.99               \\ \cline{2-7}
                       & 24        & 24.12                    & 51.98                    & 0.00142(14)  & 0.1294(33)  & 0.93               \\ \cline{2-7}
                       & 25        & 27.85                    & 51.98                    & 0.00125(32)  & 0.1355(77)  & 1.04               \\ \cline{2-7}
                       & 26        & 27.85                    & 51.98                    & 0.00205(81)  & 0.162(12)   & 1.90               \\ \hline
46.29                  & 19        & 20.39                    & 100.23                   & 0.002407(19) & 0.07201(28) & 0.75               \\ \hline
\end{tabular}
\caption{Results of the fits to the long-distance behavior of the Polyakov-loop three-point correlator $G^{(3)}(R_Y)$, described by eq.~\eqref{eq:long-distance} for various values of $\beta$ and $N_t$.}
\label{tab:highT_appendix}

\end{table}

We can fit these values of $E_0$ (together with those in tab.~\ref{tab:highT}) using the model in eq.~\eqref{eq9} (and compare to that in eq.~\eqref{eq8}), allowing us to test those model. In this process, we also extract the numerical values for the string tension in the three-point, closed channel, as the only free parameter of our fits. 

In order to have acceptable values of the $\chi^2$, we had to exclude the data point at the highest (closest to the deconfinement transition) temperature. This leave us with the data from simulations at $T < 0.8 \, T_c$. Even in this case, the linear model leads to values of the $\chi^2$ which are at least one order of magnitude larger than the number of degree of freedom of the fit. The model from eq.\eqref{eq9}, instead, leads to acceptable results for all lattice spacings, but the most rough one, as show in tab.~\ref{tab:fulNG_fit}.

\begin{table}[!htb]
    \centering
    \begin{tabular}{|l|l|l||l|}
         \hline
         $\beta$ & $\sigma \, a^2$ & $\chi^2 / N_{d.o.f}$ & $\sigma \, a^2$ open channel\\ \hline
         36.33   & 0.009051(14)    &    7.4               &  0.009022(53)               \\ \hline
         39.65   & 0.007552(12)    &    2.1               &  0.007518(23)               \\ \hline
         42.97   & 0.006375(14)    &    1.7               &  0.006378(18)               \\ \hline
         46.29   & 0.005469(9)     &    0.42              &  0.005460(27)               \\ \hline
    \end{tabular}
    \caption{Results form fit of the $E_0$ values in tab.~\ref{tab:highT} (large distance expansion only) and tab.~\ref{tab:highT_appendix}, according to the mode in eq.~\eqref{eq9} excluding the highest value of the temperature. In the last column, we report the values for the string tension in the open channel, already listed in tab.~\ref{tab:fit_res_baryon}. At least for the finer lattice spacings, the quality of the fit is satisfactory and the agreement between the two value of $\sigma$ excellent.}
    \label{tab:fulNG_fit}
\end{table}

In order to better highlight the agreement with the values obtained in the open channel, we also report the fifth column of tab.~\ref{tab:fit_res_baryon}, at the same values of $\beta$.

The deviation from the model observed in the points at higher temperature can be due to the corrections to the Nambu--Got{\=o} actions that were carefully analyzed in \cite{Caselle:2024zoh} for the two-point case. However we don't know how those correction would manifest in the three-point functions. Inserting the numerical correction determined in \cite{Caselle:2024zoh} did not improve the quality of the fit.

\FloatBarrier

\providecommand{\href}[2]{#2}\begingroup\raggedright\endgroup

\end{document}